\documentclass[times,twocolumn,final]{elsarticle}
\usepackage{medima}
\usepackage{framed,multirow}

\usepackage{amssymb}
\usepackage{latexsym}

\usepackage{url}
\usepackage{xcolor}
\usepackage{hyperref}
\usepackage{graphicx}
\usepackage[font=small]{caption}
\usepackage{amsmath,amssymb,amsfonts}
\usepackage[utf8]{inputenc}
\usepackage{booktabs, multicol, multirow}       % professional-quality tables
\usepackage{amsfonts}       % blackboard math symbols
\usepackage{nicefrac}       % compact symbols for 1/2, etc.
\usepackage{microtype}      % microtypography
\usepackage{bm}
\usepackage{amsmath,amssymb,amsfonts}
\usepackage{algorithmic}
\usepackage{graphicx}
\usepackage{float}
\usepackage{textcomp}
\usepackage{xcolor}
\usepackage{caption}
\usepackage{subcaption}
\usepackage{tipa}
\usepackage{amssymb}
\usepackage{svg}
\usepackage{caption} 
\usepackage{adjustbox}
\usepackage{comment}
\usepackage{tabularx}
\usepackage{listings}
\usepackage[ruled,vlined]{algorithm2e}
\usepackage{blindtext, graphicx}
\usepackage{listings}

\definecolor{newcolor}{rgb}{.8,.349,.1}

\journal{ArXiv}

\begin{document}

\verso{Nicol\`o Savioli \textit{et~al.}}

\begin{frontmatter}

\title{Joint Semi-supervised 3D Super-Resolution and Segmentation with Mixed Adversarial Gaussian Domain Adaptation}

%\tnotetext[tnote1]{This is an example for title footnote %coding.}

\author[1,2]{Nicol\`o Savioli}
\author[1]{Antonio de Marvao}
\author[2,3]{Wenjia Bai}
\author[2]{Shuo Wang}
\author[4]{Stuart A. Cook}
\author[4]{Calvin W. L. Chin}
\author[2,6]{Daniel Rueckert}
\author[1]{Declan P. O'Regan}

%% Third author's email
\ead{declan.oregan@imperial.ac.uk}
\address[1]{MRC London Institute of Medical Sciences, Imperial College London, London, UK}
\address[2]{Department of Computing, Imperial College London, London, UK}
\address[3]{Department of Brain Sciences, Imperial College London, London, UK}
\address[4]{National Heart Research Institute Singapore, National Heart Center Singapore, Singapore}
\address[6]{Institute for Artificial Intelligence and Informatics, Klinikum rechts der Isar, Technical University of Munich, Germany}

%\received{0000}
%\finalform{0000}
%\accepted{0000}
%\availableonline{0000}
%\communicated{S. Sarkar}

\begin{abstract}
 %%%%
 Optimising the analysis of cardiac structure and function requires accurate 3D representations of shape and motion. However, techniques such as cardiac magnetic resonance imaging are conventionally limited to acquiring contiguous cross-sectional slices with low through-plane resolution and potential inter-slice spatial misalignment. Super-resolution in medical imaging aims to increase the resolution of images but is conventionally trained on features from low resolution datasets and does not super-resolve corresponding segmentations. Here we propose a semi-supervised multi-task generative adversarial network (Gemini-GAN) that performs joint super-resolution of the images and their labels using a ground truth of high resolution 3D cines and segmentations, while an unsupervised variational adversarial mixture autoencoder (V-AMA) is used for continuous domain adaptation. Our proposed approach is extensively evaluated on two transnational multi-ethnic populations of $1,331$ and $205$ adults respectively, delivering an improvement on state-of-the-art methods in terms of Dice index, peak signal‐to‐noise ratio, and structural similarity index measure. This framework also exceeds the performance of state-of-the-art generative domain adaptation models on external validation (Dice index $0.81$ vs $0.74$ for the left ventricle). This demonstrates how joint super-resolution and segmentation, trained on 3D ground-truth data with cross-domain generalization, enables robust precision phenotyping in diverse populations.
 %%%%
\end{abstract}

\end{frontmatter}

\hfill \break

\section{Introduction}

Image super-resolution (SR) is a class of techniques in computer vision that is used to reconstruct a high resolution (HR) image from observed low resolution (LR) images. Improving the details of an image is valuable across a range of natural image applications including data compression \citep{cao2020lossless} and feature extraction \citep{rasti2016convolutional}. It is particularly relevant in medical imaging where there are physical constraints on the acquired spatial resolution and high-resolution structural detail may hold discriminative clinical information \citep{cui2014deep}. SR representations are important for cardiac imaging where detailed variations in structure and motion are informative for genetic association studies and prognostic stratification \citep{bello2019deep, schafer2017titin}. While acquiring three-dimensional cardiac magnetic resonance (CMR) cine imaging is feasible \citep{moghari2018free}, routine clinical practice and population biobanks are still dominated by multi-slice 2D cine imaging, which limits accurate assessment of geometry due to poor through-plane resolution. 

Dictionary-based approaches, such as deformable patch-match methods, aim to recover HR patches from LR images via priors between the target image and training data \citep{zhu2014single}. However, such methods are computationally demanding as the candidate patches have to be searched to find the most suitable match. In contrast, SR convolutional neural networks learn an end-to-end mapping between LR and HR images, and have significantly  improved the quality and efficiency of reconstructions \citep{gholipour2010robust,sanchez2018brain,chaudhari2018super,chen2018efficient,dong2014learning}. These methods use a convolutional neural network (CNN) that codifies non-linear transformations (i.e., patch extraction, nonlinear mapping, and reconstruction) with a combination of ResNet blocks and generative adversarial networks (GANs) \citep{ledig2017photo}. Making use of multiple images acquired from different slice directions can further improve and constrain the HR image reconstruction \citep{oktay2017anatomically,davatzikos2003hierarchical}. Lately, optical flow interpolation has been proposed for intensity cardiac image SR  \citep{xia2021super}. This method is based on recent advances in the field of video interpolation \citep{bao2019depth}, where intermediate sections are synthesized through an optical flow transformation of the cine data. However, the proposed method cannot eliminate slice misalignment and inter-slice motion artifacts (i.e., where differing breath-hold positions during scanning generate imperfect alignment between slices).  

These strategies perform segmentation independently of processing the intensity image and do not share the same feature encoder for the related tasks of super-resolving the greyscale image and its semantic label map. Multi-task deep learning approaches with atlas propagation for shape-refined segmentation can produce anatomically smooth models that are robust to the presence of artifacts in the input CMR volumes \citep{8624549}. A limitation is the computational cost needed to perform atlas propagation and the output is only HR label maps without an end-to-end super-resolved image. Joint SR and segmentation using GANs (segSR-GAN) have been proposed for brain imaging \citep{delannoy2020segsrgan} as an extension of SR-GAN \citep{ledig2017photo}. The segSR-GAN model initially performs both interpolation and patch image subdivisions of the LR input and maps it to an output within two separate up-convolutional layers. However, the initial pre-interpolation step may amplify artifacts such as the motion between successive slices or intensity bias within a given slice. Also, both pre-interpolation and/or patch image subdivisions have long computational times and may not be suitable for clinical applications or large datasets. 

In common with other classification tasks, if the training and test images for SR are drawn from different distributions, domain adaptation is needed to reduce sample bias to improve generalization performance across populations \citep{wang2018deep}. There are three main techniques for domain adaptation: divergence-based domain adaptation, adversarial-based domain adaptation, and reconstruction-based domain adaptation \citep{DBLP:journals/corr/abs-2007-15567,hoffman2018cycada,yang2020label}. The divergence-based domain adaptation method works by minimizing a divergence criterion between source and target distributions for achieving domain invariant feature representation. In adversarial-based domain adaptation, a synthetic target data domain is generated from the source data by applying GANs. A domain confusion loss is alternately applied to balance the distributions of the source and target domain in one shot within a confusion metric in the final regression layer \citep{tzeng2014deep}. Finally, in the reconstruction-based domain adaptation approach, a shared representation of the domain simultaneously solves classification and reconstruction maintaining information in the target domain \citep{ghifary2016deep}. However, the adversarial-based domain adaptation is the most used method in the medical field \citep{perone2019unsupervised,dou2018unsupervised}, but is still prone to mode collapse \citep{durall2020combating} that leads the domain adaption model to collapse in the wrong distribution. The variational adversarial autoencoder model proposed here has two main advantages: it solves the mode collapse problem by decoupling the distribution transformation within two variational networks and, at the same time, simplifies the mapping operation between source and target distributions. 

In this work, we propose: i) a generative adversarial network called Gemini-GAN that jointly performs SR of greyscale cine images and their label maps and (ii) a generative domain adaptation approach (Variational  Adversarial  Mixture  Autoencoder, V-AMA) which together with joint SR, generalizes across different populations. This approach enforces cross-domain consistency and benefits from complementary learning of super-resolving images and their segmentations. In contrast to previous work, we use a natively HR ground truth for training and validate our approach on two independent multi-ethnic cardiac MR datasets.

\hspace{2cm}

\section{Datasets}

We used paired natively acquired LR/HR images and their respective segmentations for model training, taking advantage of the near-isotropic HR 3D cine datasets available in the UK Digital Heart Project (UKDHP). We then generalized the model to an external population with equivalent paired LR/HR data available (National Heart Center Singapore Biobank), and then to a further population with only LR data available (UK Biobank) which is an intended use case. Table \ref{table:table1} summarizes the phenotypic characteristics of the datasets used. Images were stored on an open-source database (MRIdb, Imperial College London, UK) \citep{woodbridge2013mridb}. Ground truth labels for both LR and HR datasets were derived as previously described \citep{bai2015bi}. In each case ethical approval and written informed consent was obtained.

\subsection{UK Digital Heart Project (UKDHP)}

A dataset of $1331$ healthy adults was used from the UK Digital Heart Project at Imperial College London. High-spatial resolution 3D balanced steady-state free precession cine sequences were used that assessed the left and right ventricles in their entirety in a single breath-hold (60 sections, reconstructed voxel size 1.2 × 1.2 × 2 mm, 20 cardiac phases, typical breath-hold 20s). Conventional single slice multi-breath-hold images were also acquired in the same geometry (10 sections, reconstructed voxel size 1.8 × 1.8 × 8 mm). Imaging was performed on a 1.5-T Philips Achieva system (Best, the Netherlands).

\subsection{National Heart Center Singapore Biobank (SG)}

External validation was performed on a dataset of $205$ healthy adults recruited to the National Heart Center Singapore Biobank using the same LR/HR image acquisition parameters as the UKDHP but obtained on a 1.5T Aera (Siemens Healthcare, Erlangen, Germany).

\subsection{UK Biobank (UKBB)}

Generalisation to a second external dataset with only LR cine images was performed in $1331$ adults prospectively recruited to UK Biobank \citep{bycroft2018uk}. Conventional LR conventional single slice multi-breath-hold images were acquired (10 sections, reconstructed voxel size 1.8 × 1.8 × 8 mm) performed to a standard protocol on a 1.5T Aera (Siemens Healthcare, Erlangen, Germany) \citep{petersen2015uk}. Data was processed under access number 40616.

%%%%%%%%%%%%%%%%%%%%%%%%%%%%%%%%%%%%%%%%%%%%%%%%%%%%
%%% re-ordered Table 1 with SG in middle column %%%
%%%%%%%%%%%%%%%%%%%%%%%%%%%%%%%%%%%%%%%%%%%%%%%%%%%%

\begin{table}[ht]
\centering % used for centering table
 \scalebox{0.9}{
\begin{tabular}{l c c c} % centered columns (4 columns)
\hline\hline %inserts double horizontal lines
Characteristic & UKDHP & SG & UKBB$^{*}$ \\ [0.5ex] % inserts table
  & (n=1,331) & (n=205) & (n=1,331) \\
%heading
\hline % inserts single horizontal line
Age (years)         & 37.7 $\pm$ 12.51    & 50.2 $\pm$ 14.69   & 56.5 $\pm$ 8.1 \\ % inserting body of the table
Sex (male)          & 578.0  (43.4$\%$)  & 106 (51.7$\%$)   & 86.7 (54.9$\%$)  \\
Ethnicity \\
\quad Caucasian     & 922.0 (69.3$\%$)  & - & 150.4 (95.2$\%$) \\
\quad Chinese       & -                 & 191 (93.17$\%$)  & - \\
\quad South Asian   & 180  (13.5$\%$)   & 11  (5.3$\%$)    & - \\ 
\quad African       & 145  (10.9$\%$)   & -                & - \\
\quad Other         & 84    (6.3$\%$)   & 3   (1.4$\%$)    & - \\ 
BSA ($m^2$)         & 1.8$\pm$0.2       & 1.7$\pm$0.2      & 1.9$\pm$0.22 \\ 
SBP ($mmHg$)        & 118.3$\pm$14.3     & 134.4$\pm$18    & 139.7$\pm$19.6\\ 
DBP ($mmHg$)        & 77.9$\pm$9.5       & 80.1$\pm$11.6   & 82.2$\pm$10.7 \\
\hline
\end{tabular}}
\caption{Population characteristics. Values are n ($\%$) or mean $\pm$ standard deviation. $^{*}$ Characteristics shown for whole population. SG, Singapore biobank; UKDHP, UK Digital Heart Project; UKBB, UK Biobank; BSA, body surface area; SBP, systolic blood pressure; DBP, diastolic blood pressure.}
\label{table:table1} 
\end{table}

\section{Methods}

Here we describe the proposed Gemini-GAN for joint super-resolution of greyscale images and segmentation labels from routine 2D cine sequences. We then describe how V-AMA uses a novel domain transfer technique to map a source distribution to the target distribution where Gemini-GAN was previously trained. The whole framework combines both methods (V-AMA and Gemini-GAN) and enables joint SR to be generalised across domains.

%%%%%%%%%%
% Fig 09 %
%%%%%%%%%%
\begin{figure*}[t]
\includegraphics[width=\linewidth]{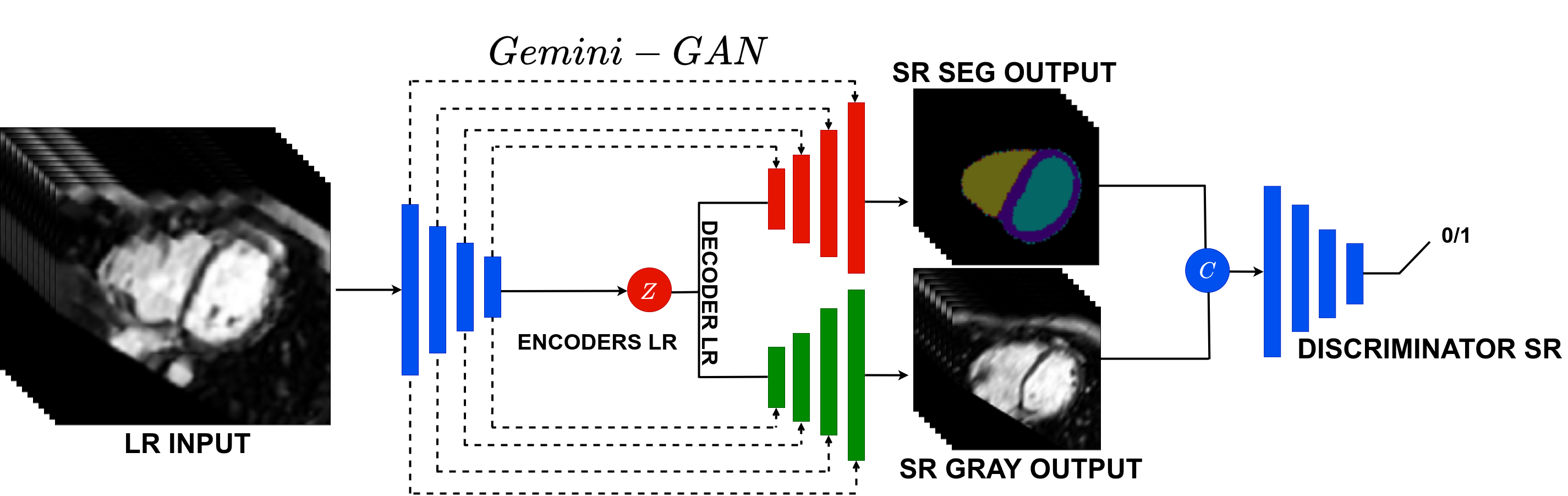}
	\centering
	\caption[\textbf{Figure 1}]{The figure shows the main components of the proposed $G_{Gemini}$-GAN model. It is constituted by a (blue) Convolution Neural Network decoder that takes as input the LR volume that is sent into the latent space $Z$ (red), which contains both SR and 3D segmentation latent information. Subsequently, the $Z$ information is sent in two encoder networks constituted by up-convolutions. The first decoder (red) reconstructs the 3D segmentation $I^{SR}_{seg}$, while the other (green) $I^{SR}_{grey}$, the volumetric SR images. The dashed lines represent the skip path connections that concatenate the features of each decoder layer with the features of the corresponding upsampling layer. Then these ``fake''-generated volumes $I^{SR}_{seg}$ and $I^{SR}_{grey}$ are both concatenated and sent to the SR discriminator to fool it into believing that they are similar to the concatenation of samples from real dataset.}
\label{fig:fig09}
\end{figure*}

\subsection{Gemini-GAN model}

The proposed Gemini-GAN (Fig. \ref{fig:fig09}) is based on the UNet style model \citep{ronneberger2015u}, and consists of one encoder path and two decoder paths (i.e., one for SR and one for segmentation reconstruction). The encoder path is composed of repeated $3 \times 3$ convolutions, followed by a Rectified Linear Unit (ReLU) and a Batch Normalisation (BN) layer. The twin's decoder branches are made by up-convolutions concatenated with the corresponding cropping feature maps from the decoder path. The networks take as input a 2D stack of LR slices $d \times w \times h$, where $d$ is the number of slices along the z-axis and $w \times h$ is the size of each cardiac slice. The input channels $d$ of the network are equal to the size of the $z$ axis's input volume. The feature maps in the latent space are split into two decoders: one for reconstructing the SR greyscale image and one for the 3D segmentation. Each feature map of each decoder layer is concatenated to the upsampling layer for both SR and 3D segmentation through skip path connections. We refer to this network as Generator $G_{Gemini}$. The energy function for the segmentation encoder branch of $G_{Gemini}$ is a cross-entropy criterion that combines log-softmax and log-likelihood. 

The log softmax [Eq. \ref{eq:1}] takes as input a 3D dimensional input. 
\begin{equation} \label{eq:1}
p_{f}(x) = \frac{e^{z_{f}(x)}}{\sum_{i=1}^{K} e^{z_{i}(x)}}
\end{equation}

Here $z_{d}(x)$ in the numerator denotes the activation map at the feature channel $f$ and pixel position $x \in \mathbb{R}^{2}$ for each cardiac slice. While the sum of the denominator is a normalization across $K$ segmentation class and ensures that the sum of probability components in the output vector $p_{f}(x)$ are equal. A Cross Entropy ($CE_{Loss}$) loss is then defined as the log-likelihood of probability of log-softmax distribution $p_{f}(x)$. The weight map $w_{f}(x)$ assigning importance to each segmentation $K$ class pixel as follows:

\begin{equation} \label{eq:2}
CE_{Loss} = w_{f}(x) \cdot log(p_{f}(x))
\end{equation}
Instead in the SR reconstruction branch a Mean Square Error (MSE) is used:
\begin{equation} \label{eq:3}
MSE_{Loss} = \frac{1}{n} \sum_{i=1,j=1}^{n} (o_{i,j} -y_{i,j})^{2}
\end{equation}

The $o_{i,j}$ is the output from the SR decoder and $y_{i,j}$ the corresponding SR ground truth, $n$ is the number of image pixels $[i,j]$.

%%%%%%%%%%%%%%%%%%%%%%%%%%%%%%%%%%%%%%%%%%%%
The L2 penalty loss is then defined as the square root of $MSE$:
%%%%%%%%%%%%%%%%%%%%%%%%%%%%%%%%%%%%%%%%%%%%
\begin{equation} \label{eq:5}
\begin{split}
L2_{Loss} =  \beta \cdot {\sqrt{(o_{i,j} -y_{i,j})^{2}}} = \beta \cdot \sqrt{nMSE_{Loss}}
\end{split}
\end{equation}
%%%%%%%%%%%%%%%%%%%%%%%%%%%%%%%%%%%%%%%%%%%%

%%%%%%%%%%%%%%%%%%%%%%%%%%%%%%%%%%%%%%%%%%%%
To increase the performance of the network an SR discriminator $D_{SR}$ is also coupled which corresponds to a network formed by eight  $3 \times 3$ reiterated convolutions followed by BN and LeakyReLU with a threshold value set at $0.2$. The generator network $G_{Gemini}$ maps the LR volume as a ``fake'' greyscale SR and 3D segmentation volumes. The concatenation operation ($+$) is applied to both ``fake''-generated volumes sampling by $I^{SR}_{fake} \sim (I^{SR}_{seg}+I^{SR}_{grey})$ where $I^{SR}_{seg}$ and $I^{SR}_{grey}$ are outputs of $G$ net. Then, the concatenation $I^{SR}_{fake}$ is sent to the SR discriminator $D_{SR}$ [Fig. \ref{fig:fig09}] trained to distinguish it from the corresponding ``real''-dataset concatenation $I^{SR}_{real} \sim (I^{SR}_{seg}+I^{SR}_{grey})$ of greyscale SR and 3D segmentation sampling from training dataset distribution. The final GAN loss is formulated as follows: 
%%%%%%%%%%%%%%%%%%%%%%%%%%%%%%%%%%%%%%%%%%%%

\begin{equation} \label{eq:6}
\begin{split}
GAN_{Loss} = \min_{G} \max_{D_{SR}} L_{GAN_{Gemini}}(D_{SR},G) = \\  = \mathbb{E}_{I^{SR}_{real} \sim UKDHP_{({I^{SR}_{seg}+I^{SR}_{grey}})}} [log(D_{SR}(I^{SR}_{real})] + \\ \mathbb{E}_{I^{SR}_{fake}} \sim  G_{({I^{SR}_{seg}+I^{SR}_{grey}})} [log(1-D_{SR}(I^{SR}_{fake})]
\end{split}
\end{equation}

%%%%%%%%%%%%%%%%%%%%%%%%%%%%%%%%%%%%%%%%%%%%
The final joint SR and 3D segmentation loss is a weighted combination of cross-entropy, MSE, GAN, and L2 losses mediate by $\lambda$, $\omega$, and $\beta$ constants:
%%%%%%%%%%%%%%%%%%%%%%%%%%%%%%%%%%%%%%%%%%%%

\begin{equation} \label{eq:7}
\begin{split}
Loss_{G_{Gemini}} = CE + \lambda \cdot MSE + \omega \cdot GAN + \beta \cdot  L2 
\end{split}
\end{equation}

%%%%%%%%%%
% Fig 10 %
%%%%%%%%%%

\begin{figure*}[t]
\includegraphics[width=1.0\linewidth]{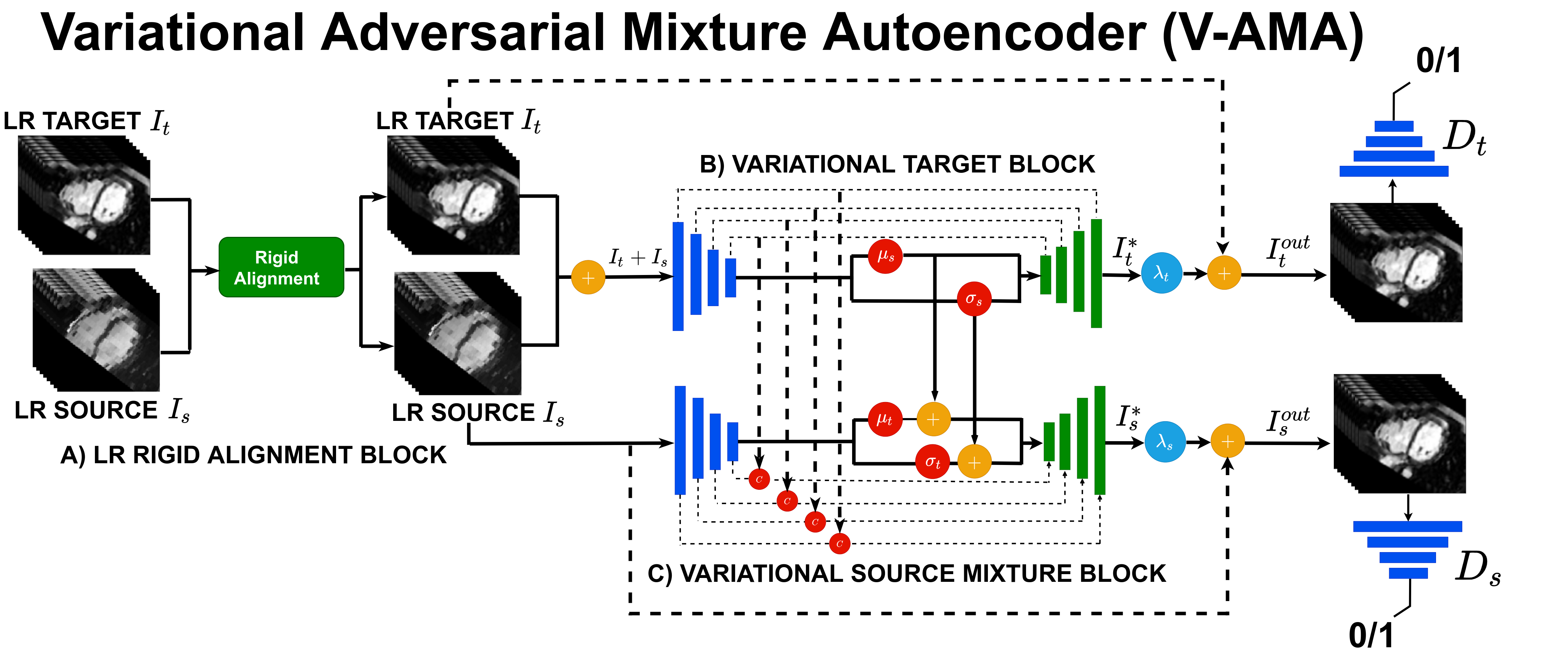}
	\centering
	\caption[\textbf{Figure 1}]{The figure shows the novel Variational Adversarial Mixture Autoencoder (V-AMA) used for the domain adaption. Particularly, the aim of the network is to adapt the LR information, from an unknown source dataset (i.e., SG, UKBB), to a target one (i.e., UKDHP dataset). The V-AMA model is composed of three distinct blocks: LR rigid alignment block, variational target block, and variational source mixture block. (a) The LR rigid alignment block allows the transformation process through an adaption network we calculated a rigid transformation among them within a set of six landmark points extracted from the apex and base of their respective ventricles. This process is fully automated as we used the respective LR segmentation both for source and target domain volumes. (b) The Variational target block takes input the concatenation of low-resolution target input and low resolution source input (i.e., $I_{i}=I_{t}+I_{s}$) and with network encoder (blue in figure) a posterior target distribution $log(q_{\phi}(z_{t}|I_{i})_{t}$ (i.e., eq \ref{eq:8}) is estimated. The image output $I_{t}^{out}$ is reconstructing from ($\mu_{t}$,$\sigma_{t}$), where the input target image $I_{t}$ is added to target output $I^{*}_{t}$. Notably, to enforce the match between the generate target image $I_{t}^{out}$ and the target volume $I_{t}$, a discriminator $D_{t}$ is used to minimize a GAN loss. c) Variational source mixture block uses the previous target posterior information ($\mu_{t}$,$\sigma_{t}$) to move the ($\mu_{s}$,$\sigma_{s}$) closer the final target distribution.}
 	\label{fig:fig10}
\end{figure*}

%%%%%%%%%%
% Fig 17 %
%%%%%%%%%%

\subsection{Variational Adversarial Mixture Autoencoder (V-AMA)}

Here we describe our domain adaption solution. A source domain is defined as $S_{d} \in \mathbb{R}^{3}$ and a target domain as $T_{d} \in \mathbb{R}^{3}$, where $S_{d}$ is a LR volume from a specific unknown LR dataset, while $T_{d}$ is the LR training set volume used by Gemini-GAN. We then train a novel network called Variational Adversarial Mixture Autoencoder (V-AMA) (Fig. \ref{fig:fig10}). The V-AMA network maps an unknown source LR input volume $I_{s} \in S_{d}$ to the training target $I_{t} \in T_{d}$ LR volume. The V-AMA is composed of three distinct blocks: LR rigid alignment block, variational target network, and variational source mixture network.  

\subsubsection{Low Resolution Rigid Alignment Block}

The LR rigid alignment block allows the alignment between LR source volume of $S_{d}$ and a target volume of $T_{d}$ through rigid transformation between them within a set of six landmark points extracted between the apex and base of the ventricles. This simplifies the adaption task; as it only needs to adapt the intensity distribution of the source domain within its target one. The rigid registration process is based on a Singular Value Decomposition (SVD) of the point locations matrix. This method determines the least-squares fit for the distance of the corresponding domain landmarks. Future developments of this block include the automatic prediction of landmarks through a regression neural network and the use of a deep learning algorithm to perform the final rigid transformation. This block is only essential in the training phase to align the target and source volumes and have a correct mapping between them. 

\subsubsection{Variational Target Network  Block}

The variational target network block is a deep latent variable generative model for estimating the log-likelihood and the posterior $S_{t}$ target distribution. The model is composed of an UNet style \citep{ronneberger2015u} encoder-decoder variational target generator network $G_{t}$, followed by a discriminator network $D_{t}$. Equivalently, for Gemini-GAN, the decoder path is constituted by $3 \times 3$ reiterated convolutions followed by a ReLU and BN. The outputs encoder is made of up-convolutions concatenated with cropped feature maps from the input decoder path. During the training, the variational target generator network $G_{t}$ (Fig. \ref{fig:fig10}), takes the concatenation of both LR source $I_{s} \in T_{s}$  and target volume $I_{t} \in T_{d}$ as a batch of 2D slices (i.e., subscript $s$ denotes the source domain and $t$ denotes the target domain).

The input to the target network (i.e., blue encoder in the variational target block \ref{fig:fig10}) is indicated as a concatenation operation (+) between the target volume $I_{t}$ and source volume $I_{s}$ to better learn the consistency between both (i.e.,  $I_{i}=I_{t}+I_{s}$). After the encoding process  $I_{i}$ is mapped to a Gaussian posterior distribution in the latent space $q_{\phi}(z_{t}|I_{i})_{t} = \mathcal{N}(z_{t};\mu_{t},diag(\sigma_{t}))$. In particular, $\phi$ are the variational parameters, $z_{t}$ the latent variable of the target $t$ encoder, and $I_{t} \in T_{d}$ the LR target volume. The following equations explain the variational target encoding (i.e., please refer to theorem \ref{appendix:Theorem1} for more precise mathematical explanation): 

\begin{equation} \label{eq:8}
\begin{split}
I_{i}=I_{t}+I_{s} \\
(\mu_{t}, log (\sigma_{t})) = Encoding_{{G_{t}}_{\phi}}(I_{i}) \\
q_{\phi}(z_{t}|I_{i})_{t} = \mathcal{N}(z_{t};\mu_{t},diag(\sigma_{t})) \\
log(q_{\phi}(z_{t}|I_{i})_{t}) = \sum_{k} log(\mathcal{N}(z_{{t}_{k}},0,1)) - log (\sigma_{{t}_{k}}) \\
= \sum_{k} \frac{1}{2} (log(\sigma_{t_{k}}) - \mu^{2}_{t_{k}} - \sigma^{2}_{t_{k}}+1)
\end{split}
\end{equation}

The reconstructed target output $ I^{*}_{t}$ (i.e., Fig. \ref{fig:fig10}) is concatenated with $I_{t}$ in order to constrain the match between the $I_{i}$ and $I_{t}$. The reconstruction target $I^{out}_{t}$ (in output from the variational target network) is given by:

\begin{equation} \label{eq:20}
I^{out}_{t} = I_{t} + \lambda_{t} \dot I^{*}_{t}  
\end{equation}

where $\lambda_{t}$ constant mediates the concatenation between $I_{t}$ and $I^{*}_{t}$. Also, to better enforce the match between the generated target volume $I^{out}_{t}$ and the input target volume $I_{t}$ (i.e., sample from the UKDHP dataset), a discriminator $D_{t}$ is also used to minimize the following GAN loss for the variational target generator network $G_{{t}_{\phi}}$ (i.e., please also see theorem \ref{appendix:Theorem2}):

\begin{equation} \label{eq:21}
\begin{split}
GAN^{t}_{Loss}= \min_{G_{{t}_{\phi}}} \max_{D_{t}} L_{GAN_{t}}(D_{t},G_{{t}_{\phi}}) = \\  = \mathbb{E}_{I^{real}_{t} \sim {UKDHP(I_{t})}} [log(D_{t}(I^{t}_{real}))] + \\ \mathbb{E}_{I^{fake}_{t} \sim G_{{t}_{\phi}}(I_{i})} [log(1-D_{t}(I^{t}_{fake})] 
\end{split}
\end{equation}

Then, the total target loss [Eq. \ref{eq:27}], for the variational target network block, is composed by the sum of ELBO loss (i.e., Eq \ref{eq:15} in the appendix), GAN loss [Eq. \ref{eq:21}] and L1 loss. If we rearranging [Eq. \ref{eq:13}, \ref{eq:14}, and \ref{eq:15}] we have: 

\begin{equation} \label{eq:26}
\begin{split}
VAELoss_{\theta,\phi} (I_{i}) = log(p_{\theta}(I_{i})) - log(q_{\phi}(z_{t}|I_{i}) + p_{\theta}(z_{t}|I_{i}) \\ 
=  log(p_{\theta}(I_{i}) \cdot p_{\theta}(z_{t}|I_{i})) -  log(q_{\phi}(z_{t}|I_{i}) \\
=  p_{\theta}(I_{i},z_{t}) - log(q_{\phi}(z_{t}|I_{i})) \\
\end{split}
\end{equation}

The $p_{\theta}(I_{i},z_{t})$ distribution is approximated by a MSE loss between reconstruction target $I^{out}_{t}$ (i.e., Eq. \ref{eq:20}) and variational network input (i.e., $I_{i}=I_{t}+I_{s}$). While the posterior distribution $q_{\phi}(z_{t}|I_{i})$ is given by the [Eq. \ref{eq:80}] in the appendix. 

\begin{equation} \label{eq:27}
\begin{split}
VAELoss_{\theta,\phi} (I_{i})_{t} = MSE(I_{i},I^{out}_{t}) + \\ 
0.5 \cdot \sum_{i} \frac{1}{2}(\sigma^{2}_{t_{i}} + \mu^{2}_{t_{i}} - log(\sigma_{t_{i}})-1)
\end{split}
\end{equation}

Final target loss (i.e.,  variational target block in Fig. \ref{fig:fig10}) is: 

\begin{equation} \label{eq:28}
\begin{split}
Target_{Loss} = VAELoss_{\theta,\phi} (I_{i}) + \\ L1_{Loss}(I_{i},I^{out}_{t}) + 0.05*GAN^{t}_{Loss}
\end{split}
\end{equation}

\subsubsection{Variational Source Mixture Network Block}

After training the variational target block, both encoding features (i.e., at each variational encoding layers) and inference $(\mu_{t}$ and $\sigma_{t} )$ [Eq. \ref{eq:8}] are transferred to the source encoder  (i.e., indicated with $G_{{s}_{\phi}}$) to equally inject the same amount of target distribution (Fig.\ref{fig:fig10}). The posterior distribution $q_{\phi}(z_{s}|I_{s})_{s}$ (i.e., $s$ in the following equations represent the source domain $T_{s}$) is derived by: 

\begin{equation} \label{eq:29}
\begin{split}
(\mu_{s}, log (\sigma_{s})) = Encoding_{{G_{s}}_{\phi}}(I_{s}) \\
q_{\phi}(z_{s}|I_{s})_{s} = \mathcal{N}(z_{s};\mu_{s},diag(\sigma_{s})) \\
log(q_{\phi}(z_{s}|I_{s})_{s}) = \sum_{k} log(\mathcal{N}(z_{{s}_{k}},0,1)) - log (\sigma_{{s}_{k}}) \\
= \sum_{k} \frac{1}{2} (log(\sigma_{s_{k}}) - \mu^{2}_{s_{k}} - \sigma^{2}_{s_{k}}+1)
\end{split}
\end{equation}

where the previous posterior inference of target network $q_{{\phi}(z|I_{s})_{t}}$ and the new posterior inference of source network $q_{{\phi}(z_{s}|I_{s})_{s}}$ are added together in the latent space of $G_{{s}_{\phi}}$  to constitute the final mixture KL posterior $q_{{\phi}(z_{s}|I_{s})_{m}}$. Indeed, the source network $G_{{s}_{\phi}}$ loss is:

\begin{equation} \label{eq:30}
\begin{split}
VAELoss_{\theta,\phi} (I_{s}) =  MSE(I_{s},I^{out}_{s}) + \\  0.5 \cdot \sum_{k} \frac{1}{2} (log(\sigma_{{m}_{k}}) - \mu^{2}_{m_{k}} - \sigma^{2}_{m_{k}}+1)
\end{split}
\end{equation}

Given that $(\sigma_{t},\mu_{t})$ and $(\sigma_{s},\mu_{s})$ are normally distributed independent random variables, then their sum is also normally distributed. Hence, we obtain $\sigma_{{m}^{2}_{k}}$, $log(\sigma_{{m}_{k})}$, $\mu_{{m}^{2}_{k}}$, and $z_{{m}_{k}}$ as:

\begin{equation} \label{eq:31}
\begin{split}
\sigma_{{m}^{2}_{k}} = \sigma_{t_{k}}^{2} + \sigma_{s_{k}}^{2} \\
log(\sigma_{{m}_{k}}) = log(\sigma_{t_{k}}) + log(\sigma_{s_{k}}) \\
\mu_{{m}^{2}_{k}} = (\mu_{t_{k}} + \mu_{s_{k}})^2 \\
k_{{m}_{k}} \sim  \mathcal{N}(0,I) \\
z_{{m}_{k}} = (\mu_{t_{k}}+\mu_{s_{k}}) + (\sigma_{s_{k}}+\sigma_{t_{k}}) \odot k_{m_{k}} \\
\end{split}
\end{equation}

The reconstruction target $I^{out}_{s}$ in output from $G_{{s}_{\phi}}$ is given by:

\begin{equation} \label{eq:32}
I^{out}_{s} = I_{s} + \lambda_{s} \dot I^{*}_{s}  
\end{equation}

Where $\lambda_{s}$ constant mediates the interaction between the source network input $I_{s}$ and the output $I^{*}_{s}$. As for $G_{{t}_{\phi}}$, to better enforce the match between the generated source volume $I^{out}_{s}$ and the target volume distribution $I_{t}$ (i.e., sample from the UKDHP dataset), we used a discriminator $D_{s}$ to minimize the following GAN loss for the $G_{{s}_{\phi}}$ network:

\begin{equation} \label{eq:33}
\begin{split}
GAN^{s}_{Loss}= min_{G_{{t}_{\phi}}} max_{D_{s}} L_{GAN_{s}}(D_{s},G_{{s}_{\phi}}) = \\  = \mathbb{E}_{I^{real}_{t} \sim {UKDHP(I_{t})}} [log(D_{s}(I^{t}_{real}))] + \\ \mathbb{E}_{{I^{fake}_{s}} \sim G_{{s}_{\phi}}(I_{s})} [log(1-D_{s}(I^{s}_{fake})] 
\end{split}
\end{equation}

The total source loss, for the variational source network block, is composed by the sum of posterior mixture KL loss (i.e., eq \ref{eq:30} and \ref{eq:31}), GAN loss [Eq \ref{eq:33}] and L1 loss. 

\begin{equation} \label{eq:34}
\begin{split}
Soruce_{Loss} = VAELoss_{\theta,\phi} (I_{s})  + \\ L1_{Loss}(I_{s},I^{out}_{s}) + 0.05*GAN^{s}_{Loss}
\end{split}
\end{equation}

The variational source mixture network uses the target posterior information to fully incorporate the target distribution in source one so that the source reconstruction $I^{out}_{s}$ will look like the target image UKDHP style.  The obvious benefit is linked to a better inference task in the domain adaption process where the variational target network is only involved to find the posterior distribution that captures the UKDHP style. 
However, in order to optimize the inference speed, the LR rigid alignment block is used only during the training stage where during test inference both variational target and variational source mixture networks take in input the source $I_{s}$.

\subsection{Training details}

The number of UKDHP patients utilized for training is 1131 with  $100$ epochs iterations until convergence and finally valid and testing in 100 UKDHP's patients. Data argumentation was also applied through the extraction of random cropping, within vertical and horizontal flip; where the final inference is performed for all networks comparison on the entire volume at LR. The learning rate (lr) used is $1e-4$ with a weight decay (i.e., L2 penalty) of $\beta$ set to $1e-6$ [Eq \ref{eq:5}]. Also the $\lambda=1$, for giving equal importance to SR than 3D segmentation during the training; while the GAN constant is set to $\omega=1e-3$. The LR images, and their corresponding segmentation and SR images, are used to train the network in a supervised way with ADAM optimization with random horizontal and vertical flipping and random cropping \citep{kingma2014adam}. While for V-AMA an ADAM optimization is also used with an lr of $0.0002$ and a momentum (i.e., $beta_{1}$) of 0.5 where $\lambda_{t}=1$ and $\lambda_{s}=1$. All networks were consistently compared in inference using the full cardiac LR-volume (i.e., no initial cropping and/or interpolation) inputs where the training hyperparameters were kept constant for all training phases. The GPU used is an NVIDIA V100 with a smaller batch size for minimizing the overhead between the GPU and the Hard Disk Drive (HDD) and maximizing the use of the overall deep-learning system.

\section{Results}

First we describe the metrics used to quantify algorithm performance. We then compare the performance of Gemini-GAN (i.e., super-resolution performances) to several benchmark algorithms used for either sequential or joint SR and segmentation tasks, and then evaluate V-AMA (i.e., domain-adaption performances) in an external dataset against other domain adaptation methods.  

\subsection{Segmentation and image quality metrics} 

\subsubsection{Dice Index (DI)}

Segmentation performance is evaluated for the left ventricular (LV) cavity, LV myocardium and right ventricular (RV) cavity both in end-systole (ED) and end-diastole (ED) with the Dice Index (DI). Given two sets of binary mask X and Y, the DI is defined by the following equation: 
\begin{equation} \label{eq:dice}
Dice = \frac{2|X \displaystyle \cap  Y|}{|A| + |B|}
\end{equation}
Here $|X|$ and $|Y|$ represent the number of elements for X and Y. 

\subsubsection{Peak Signal-to-Noise Ratio (PSNR)}

The PSNR represents the ratio between the maximum possible power of a signal and the power of corrupting noise that affects the fidelity of its representation. The mathematical definition of PSNR is expressed logarithmically. Given two images $I$ and $K$ of size $m \times n$ we define PSNR (in decibel scale) as: 
\begin{equation} \label{eq:psnr}
\begin{split}
PSNR = 10 \cdot log_{10}(\frac{I^{2}_{Max}}{MSE_{I,K}}) = \\ 20 \cdot log_{10}(\frac{I_{Max}}{\sqrt{MSE_{I,K}}}) = 20 \cdot log_{10} (I_{Max})  \\
- 10 \cdot log_{10} (MSE_{I,K})
\end{split}
\end{equation}
Here $I_{Max}$ is the maximum possible value of the image $I$, while $MSE_{I,K}$ is the Mean Squared Error between the image $I$ and $K$.
\begin{equation} \label{eq:ssim}
MSE_{I,K} = \frac{1}{m \cdot n} \sum^{m-1}_{i=0} \sum^{n-1}_{j=0} [I(i,j) -K(i,j)]^{2}
\end{equation}

\subsubsection{Structural Similarity Index Measure (SSIM)}

The structural similarity index measure (SSIM) is a metric for the assessment of the visual quality of images. Structural information measures the pixel inter-dependencies that emphasize visual structure and take into consideration contrast and luminance \citep{wang2004image}. Given two images x and y of size $m \times n$ the SSIM index is defined as: 
\begin{equation} \label{eq:dice}
SSIM(x,y) = \frac{(2 \mu_{x} \mu_{y} + c_{1}) \cdot (2 \sigma_{xy} + c_{2})}{(\mu^{2}_{x} + \mu^{2}_{y} + c_{1})  \cdot (\sigma^{2}_{x} + \sigma^{2}_{y} +c_{2})}
\end{equation}
Here $\mu_{x}$ is the average of image x, $\mu_{y}$ is the average of image y,$\sigma^{2}_{x}$ is the variance of image x, $\sigma^{2}_{y}$ is the variance of image y, and $\sigma_{xy}$ the covariance of image x and y. The constants $c_{1}$ and $c_{2}$ stabilize the division in the case of a weak denominator.

\subsection{Deep-learning methods comparison}

We used three state-of-the-art methods for both SR and segmentation comparison: UNet, SR-GAN, and SegSR-GAN. The UNet model is a advanced segmentation model \citep{ronneberger2015u} based on an encoding path, that captures the general context of the input image, and a decoding path that produces the final segmentation. While the SR-GAN is based on two networks a ResNet \cite{DBLP:journals/corr/HeZRS15} generator with a GAN discriminator (i.e., similar to Gemini-GAN). Lastly, the SegSR-GAN \citep{delannoy2020segsrgan} model, a leading-edge joint segmentation method, is further compared. The SegSR-GAN is essentially similar to SR-GAN in architecture with the exception of having an upsampling block that gives both segmentation and SR outputs. However, to be fully consistent with other model's comparisons (i.e., UNet, SR-GAN, and Gemini-GAN), no initial interpolation and/or image patch extraction was performed to the input (i.e., as in the original paper the SegSR-GAN). We instead prefer to respect the initial architecture proposed by SR-GAN  (i.e., adding a skip connection from the first convolution block to the last before the upsampling layer).

\subsection{Performance for Super-resolution}

\begin{figure}[htbp]
\includegraphics[width=\linewidth]{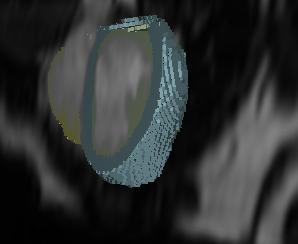}
	\centering
	\caption[\textbf{Figure 1}]{Joint 3D super-resolution and segmentation of the left and right ventricles cavity in long axis cross-section using the Gemini-GAN network showing the consistency between super-resolved greyscale images and labels prediction.}
\label{fig:fig01}
\end{figure}

\begin{figure*}[htbp]
\includegraphics[width=0.7\linewidth]{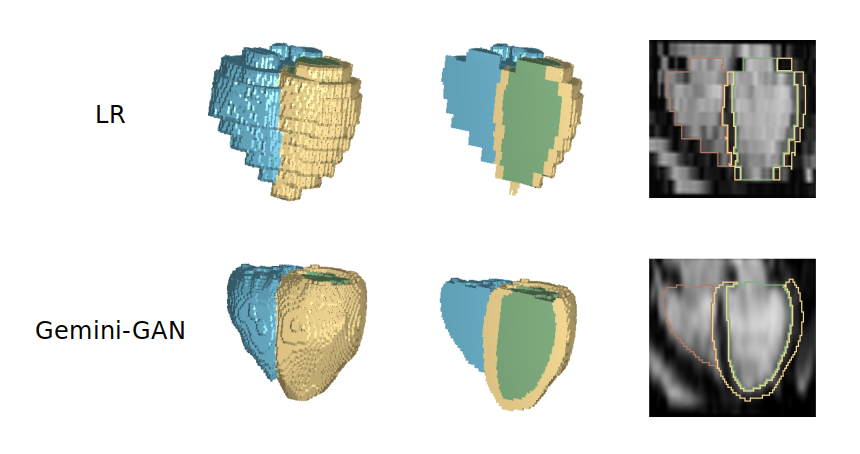}
	\centering
	\caption[\textbf{Figure 3}]{Three dimensional renderings of the surface and long-axis of the heart (yellow: left ventricular myocardium; green: left ventricular cavity; blue: right ventricular cavity) with corresponding greyscale images and segmentation contours in a long axis reconstruction. A comparison is shown between the acquired low resolution (LR) images and Gemini-GAN's joint 3D segmentation and super-resolution.}
\label{fig:fig03}
\end{figure*}

We evaluate the super-resolution performances through two main methods: (i) classic interpolation-based methods, (ii)  deep learning approaches in the UKDHP dataset which contains paired LR and HR ground truth (Table \ref{tab:sr_table}). An example of the output from Gemini-GAN is shown in Figs. \ref{fig:fig01} and \ref{fig:fig03}. The first three classic interpolation-based methods models evaluated were nearest neighbor (NN), linear, and B-spline which are all sub-optimal both in terms of 3D Segmentation and SR (and they cannot be used for an initial pre-interpolation as in SegSR-GAN paper). The performance of UNet \citep{ronneberger2015u} for SR and segmentation shows good performance in either task individually.  
We also show that the SegSR-GAN joint algorithm under-performs the proposed Gemini-GAN model which generally has the highest segmentation accuracy across all comparisons while achieving comparable super-resolved image quality to disjoint methods. As segSR-GAN is a direct derivation of SR-GAN we hypothesize that the decrease in performance for both networks is linked to two main causes: firstly the number of parameters increases with respect to the size of our dataset especially when the ResNet \citep{DBLP:journals/corr/HeZRS15} backbone is used. Secondly, the lack of skip-path connections in the upsampling layers causes a loss of high-level information responsible for shape and object detection. The number of ResNet layers utilized does not decisively influence the network performance (i.e., since an data argumentation mechanism, allows to prevent overfitting due to extra network parameters).

\begin{table*}[]
\begin{adjustbox}{width=\textwidth}
\begin{tabular}{|l|l|l|l|l|l|l|l|l|l|l|}
\hline
\textbf{Metric}      &\multicolumn{6}{c|}{\textbf{Dice}}     &\multicolumn{2}{c|}{\textbf{PSNR}}    &\multicolumn{2}{c|}{\textbf{SSIM}}          \\ \hline
\textbf{Anatomy}                 & {\color[HTML]{000000} \textbf{LV cavity	 ED}} & {\color[HTML]{000000} \textbf{LV cavity	 ES}} & {\color[HTML]{343434} \textbf{LV myocardium ED}} & {\color[HTML]{000000} \textbf{LV myocardium ES}} & \textbf{RV cavity ED}      & \textbf{RV cavity ES}      & \textbf{ED}          & \textbf{ES}          & \textbf{ED}         & \textbf{ES}         \\ \hline
\textbf{Linear}                  & 0.02(0.01)                              & 0.01(0.01)                              & 0.003(0.002)                           & 0.004(0.006)                           & 0.001(0.002)        & 0.0008(0.004)       & 12.69(0.50)          & 12.57(0.46)          & 0.46(0.03)          & 0.44(0.03)          \\ \hline
\textbf{Nearest neighbour}                      & 0.02(0.01)                              & 0.02(0.009)                             & 0.004(0.002)                           & 0.005(0.006)                           & 0.001(0.002)        & 0.0009(0.004)       & 12.68(0.50)          & 12.55(0.46)          & 0.46(0.03)          & 0.44(0.03)          \\ \hline
\textbf{B-Spline}                & 0.001(0.00)                             & 0.001(0.00)                             & 3.24e-05(0.00)                         & 9.93e-06(0.00)                         & 3.84e-05(0.00)      & 3.52e-05(0.00)      & 13.39(0.52)          & 13.41(0.48)          & 0.46(0.03)          & 0.44(0.03)          \\ \hline
\textbf{UNet \citep{ronneberger2015u}  (Only Seg)}        & 0.84(0.05)                              & \textbf{0.72(0.04)}                     & 0.72(0.04)                             & 0.72(0.06)                             & 0.71(0.05)          & 0.60(0.06)          & -                    & -                    & -                   & -                   \\ \hline
\textbf{UNet \citep{ronneberger2015u} (Only Img)}         & -                                       & -                                       & -                                      & -                                      & -                   & -                   & \textbf{27.86(0.65)} & \textbf{28.12(0.71)} & \textbf{0.90(0.01)} & \textbf{0.91(0.01)} \\ \hline
\textbf{SR-GAN \citep{ledig2017photo} (Only Seg)}       & 0.85(0.05)                              & 0.50(0.05)                              & 0.70(0.04)                             & 0.58(0.05)                             & 0.71(0.05)          & 0.59(0.07)          & -                    & -                    & -                   & -                   \\ \hline
\textbf{SR-GAN \citep{ledig2017photo} (Only Img)}       & -                                       & -                                       & -                                      & -                                      & -                   & -                   & 23.66(0.37)          & 23.53(0.38)          & 0.77(0.01)          & 0.77(0.01)          \\ \hline
\textbf{SegSR-GAN \citep{delannoy2020segsrgan}  (Joint)}      & 0.82(0.05)                              & 0.70(0.05)                              & 0.70(0.04)                             & 0.71(0.04)                             & 0.72(0.06)          & 0.58(0.08)          & 19.90(0.36)          & 20.10(0.29)          & 0.53(0.02)          & 0.54(0.01)          \\ \hline
\textbf{Gemini-GAN {[}Our{]}  (Joint)} & \textbf{0.87(0.05)}                     & 0.69(0.06)                              & \textbf{0.75(0.03)}                    & \textbf{0.75(0.04)}                    & \textbf{0.74(0.05)} & \textbf{0.64(0.05)} & 26.93(0.59)          & 27.12(0.60)          & 0.88(0.01)          & 0.88(0.01)          \\ \hline
\end{tabular}
 \end{adjustbox}
\caption{A comparison of each super-resolution model showing segmentation accuracy for each cardiac structure (Dice) and greyscale image quality metrics (PSNR and SSIM) with mean and standard deviation (Left ventricle, LV; right ventricle, RV; end diastole, ED; end systole; ES).}
\label{tab:sr_table} 
\end{table*}

%%%%%%%%%%%%%%%
% Plot Gemini %
%%%%%%%%%%%%%%%

\begin{table*}[!htbp]
\begin{adjustbox}{width=\textwidth}
\begin{tabular}{|l|l|l|l|l|l|l|l|l|l|l|}
\hline
\textbf{Metric}      &\multicolumn{6}{c|}{\textbf{Dice}}     &\multicolumn{2}{c|}{\textbf{PSNR}}    &\multicolumn{2}{c|}{\textbf{SSIM}}          \\ \hline
\textbf{Anatomy}              & {\color[HTML]{000000} \textbf{LV cavity	 ED}} & {\color[HTML]{000000} \textbf{LV cavity	 ES}} & {\color[HTML]{343434} \textbf{LV myocardium ED}} & {\color[HTML]{000000} \textbf{LV myocardium ES}} & \textbf{RV cavity ED}      & \textbf{RV cavity ES}      & \textbf{ED}          & \textbf{ES}          & \textbf{ED}         & \textbf{ES}         \\ \hline
\textbf{No Adaptation} & 0.01(0.01)                              & 0.01(0.01)                              & 0.01(0.01)                             & 0.01(0.01)                             & 0.01(0.03)          & 0.01(0.02)          & 10.68(0.71)          & 10.55(0.71)          & 0.14(0.09)         & 0.13(0.10)          \\ \hline
\textbf{CycleGAN \citep{zhu2017unpaired}}             & 0.72(0.06)                              & 0.31(0.06)                              & 0.67(0.04)                             & 0.45(0.05)                             & 0.63(0.06)          & 0.33(0.10)          & 23.35(0.46)          & 23.50(0.43)          & 0.49(0.02)          & 0.48(0.02)          \\ \hline
\textbf{BicycleGAN \citep{zhu2017toward}}           & 0.48(0.06)                              & 0.09(0.03)                              & 0.54(0.04)                             & 0.16(0.04)                             & 0.50(0.08)          & 0.14(0.07)          & 22.21(1.13)          & 21.67(1.26)          & 0.49(0.02)          & 0.48(0.02)          \\ \hline
\textbf{MUNIT \citep{huang2018multimodal}}                & 0.74(0.06)                              & 0.26(0.06)                              & 0.66(0.04)                             & 0.41(0.05)                             & 0.62(0.06)          & 0.32(0.10)          & 23.12(0.44)          & 22.87(0.42)          & 0.49(0.02)          & 0.48(0.02)          \\ \hline
\textbf{V-AMA {[}Our{]}}      &\textbf{0.81(0.05)}                              & \textbf{0.40(0.06)}                              & \textbf{0.71(0.04)}                    &\textbf{ 0.53(0.06)}                            & \textbf{0.64(0.04)} & \textbf{0.41(0.09)}          & \textbf{25.05(0.52)} & \textbf{24.80(0.54)} & \textbf{0.49(0.02)} & \textbf{0.48(0.02)} \\ \hline
\end{tabular}
\end{adjustbox}
\caption{A comparison in the external validation set (SG) between the presented V-AMA model and state of art networks (i.e., CycleGAN \citep{zhu2017unpaired}, BicycleGAN \citep{zhu2017toward}, and MUNIT \citep{huang2018multimodal}) used for image domain adaption relative to ``No Adaptation''. Segmentation accuracy for each cardiac structure (Dice) and greyscale image quality metrics (PSNR and SSIM) with mean and standard deviation are shown. (Left ventricle, LV; right ventricle, RV; end diastole, ED; end systole; ES).}
\label{tab:dmtableSG} 
\end{table*}

\begin{figure*}[p!]
\includegraphics[width=0.7\linewidth]{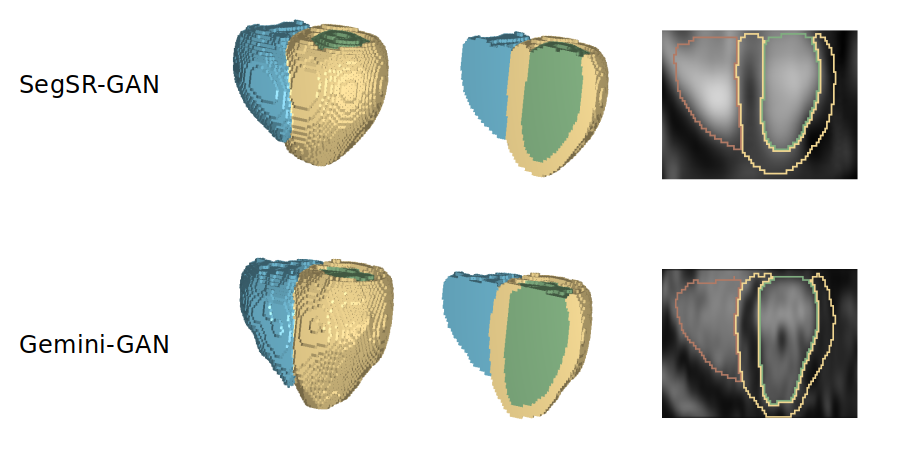}
	\centering
	\caption[\textbf{Figure 10}]{Three dimensional renderings of the heart comparing Gemini-GAN with the current state-of-the-art for joint segmentation and SR (segSR-GAN \citep{delannoy2020segsrgan}) which does not accurately reconstruct fine anatomic details in the greyscale image.}
\label{fig:fig18}
\end{figure*}

%%%%%%%%%%
% Fig 14 %
%%%%%%%%%%

\begin{figure*}[p!]
\includegraphics[width=0.8\linewidth]{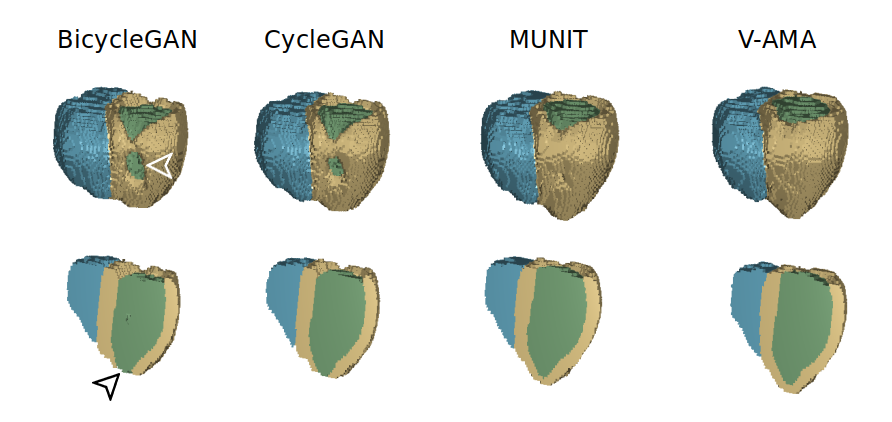}
	\centering
	\caption[\textbf{Figure 7}]{Three dimensional renderings of the heart comparing 3D segmentations produced by Gemini-GAN after adaptation with different state-of-the-art methods and our proposed V-AMA method. Our proposed framework maintains a complete and smooth geometry of the heart while other methods show discontinuities in the segmentation of the mid-ventricular and apical myocardium (arrowheads).}
 	\label{fig:fig14}
\end{figure*}

%\begin{figure}[h!]
%\includegraphics[width=0.9\linewidth]{imgs/fig_17.png}
%	\centering
%	\caption[\textbf{Figure 7}]{The figure shows the prediction of SR through our semi-supervised approach based on the combination of V-AMA and Gemini-GAN. In particular, the proactive combination of V-AMA (unsupervised learning) and Gemini-GAN (supervised learning) allows generalizing well in small datasets such as SG where SR details coincide with ground truth.}
% 	\label{fig:fig17}
%\end{figure}

%%%%%%%%%%
% Fig 20 %
%%%%%%%%%%

\begin{figure*}[p!]
\includegraphics[width=0.8\linewidth]{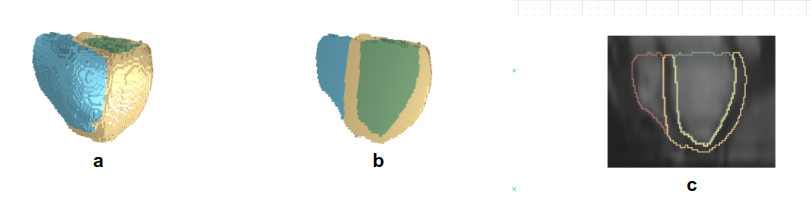}
	\centering
	\caption[\textbf{Figure 7}]{Three dimensional renderings of the heart comparing demonstrating that our Gemini-GAN + V-AMA framework generalises well to a further external dataset where only LR images are available (UKBB) demonstrating its potential value for routine clinical imaging and biobank samples. a) The full volume reconstruction from the 3D segmentation prediction, b) the volume in cross-section, and c) the greyscale SR with the 3D segmentation contours.}
 	\label{fig:fig19}
\end{figure*}

\subsection{Performance for domain adaptation}

Here we present the performance of the V-AMA network in the external validation (SG) dataset. Three state-of-the-art networks (CycleGAN \citep{zhu2017unpaired}, BicycleGAN \citep{zhu2017toward}, and MUNIT \citep{huang2018multimodal}) used for image domain adaption are compared to V-AMA (Table \ref{tab:dmtableSG} and Fig.\ref{fig:fig14}). 

The main goal of CycleGAN is to learn the mapping with a generator from a source distribution to a target one by an adversarial and a cycle consistent loss that boosts the mapping invertibility \citep{zhu2017unpaired}. To solve this problem, we use BicycleGAN - where two networks are trained with cycle loss and random noise injection.  The first network is a conditional variational auto-encoder, that encodes the target image in the latent target space and uses it to map the target image (i.e., from the source image to latent target space to target image). The second network is a conditional latent regressor network and uses the source image and random noise to produce the latent target image \citep{zhu2017toward}. The injection of random noise in the second BicycleGAN network could potentially lead to instability and is reflected by a performance decrease for the Dice index.

In the MUNIT model, the latent space source volume input is decomposed into a style space and content space. This is performed by content encoder and style encoder networks. While the style encoder network is composed of several stride convolutions, followed by global pooling and MPL layer to produce a set of parameters. The content encoder network processes the source volume to give as input to the decoding network with adjusted parameters by the style encoder. However, the style of the content encoder network is limited by the previous global pooling operation in the style encoder that is insufficient to extract detailed information due to the intrinsic limitation of pooling operation \citep{46351} (i.e., pooling takes the neuron with high peak information but not the one relevant for the specific encoding task to performed).

To address this problem, V-AMA is proposed to decouple the source latent space through two adversarial variational networks without any pooling in the style content-encoding where the variational mechanism is instead used to share style information. In the external validation set, this provided the highest segmentation accuracy for all cardiac structures as well as the best image quality metrics.

We further tested V-AMA in the UKBB dataset (which is representative of routine imaging and population data), but since there is no specific SR ground-truth both in 3D segmentation and greyscale in UKBB, we limit our analysis to a qualitative evaluation; confirming the system's ability to generalize well to a further external dataset (Fig \ref{fig:fig19}).

\section{Discussion and Conclusion}

The proposed deep-learning system creates a hybrid between supervised and unsupervised learning algorithms for training the Gemini-GAN and V-AMA algorithms respectively. We have shown the improved performance of Gemini-GAN compared to best in class networks for jointly determining 3D Segmentation and SR from a natively LR input volume while maintaining high PSNR and SSIM.

We compared our network with SR-GAN which showed relatively poor performance for segmentation accuracy and image quality. This was primarily due to a lack of a large dataset (i.e, 350,000 ImageNet images were originally used to train SR-GAN) to satisfy the depth of the ResNet networks. Furthermore, the lack of skip path connections in both SR-GAN  and Seg-SRGAN reduces the segmentation performance, as the local information in the high-level features in the first convolutions layers is lost. A further limitation of Seg-SRGAN, is related to the initial input B-spline interpolation which can introduce a prediction bias. Furthermore, interpolating the volume with a classical method, as in the original Seg-SRGAN paper, may be computationally ineffective. This can be unfavorable if we additionally want to determine the spatial information from a full volumetric LR input, where erasing of the skip path, at each up-sampling level, reduces the possibility of reaching the correct 3D spatial information between the individual channels at each encoding layer. However, Gemini-GAN exhibits high performance for both 3D segmentation and SR directly from LR greyscale volumes even with limited dataset size. 

Our unsupervised domain adaptation algorithm, V-AMA, is also proposed here. After spatial alignment of the source and target volume within a rigid alignment algorithm, paired distinct adversarial VAE networks are utilized. The first network, called variation target block, extracts the variational source volume parameters in terms of mean and std, transferred to a second network, a variational source mixture block, which mixes both source and target VAE's parameters. The V-AMA model has several advantages, in the first place it solves the mode collapse problem \citep{durall2020combating}, where the optimization of multidimensional non-convex space can lead to instability and collapse the model to a single distribution (i.e., only source but not target). Secondly, it simplifies the mapping operation between source to target distributions by disassociating the inference process within two distinct deep networks. To quantify the ability of V-AMA to adapt its domain in a small dataset we used an external dataset and three modern domain adaption networks demonstrating its accurate and robust performance.

 Mode collapse is one of the main problems for other algorithms, which could be partially solved by injecting input noise into the network as in BicycleGAN \citep{zhu2017toward}. The direct injection of random noise in the network could lead to instability especially when the source distribution differs from the target one. Notably, any additional random noise could amplify this behavior. Instead, the MUNIT model decomposes the latent space into style space and content space within two networks decreasing the mode collapse problem. Notably, one of the main limitations of MUNIT is related to global pooling in style encoding, where the global pooling layer loses relevant information and neglecting the spatial relation between the objects in the volume (i.e., LV vs RV cavity position in the volume input) \citep{sabour2017dynamic}. Instead, our V-AMA model avoids the pooling by mixing the source and target style directly in variational inference fashion.
 
In summary, we have experimentally demonstrated how our deep learning system can achieve high performance in joint cardiac SR and 3D segmentation tasks with domain adaption to new distributions. Our method will enable future applications in the field of cardiac shape and motion analysis where precision phenotypes are needed for prediction and classification tasks across diverse datasets.

% Activate the appendix
% from now on sections are numerated with capital letters

\section*{Data availability}

The code \citep{github} and segmentations \citep{data} used in this study are open access.  

\section*{Acknowledgments}
The study was supported by Bayer AG; Medical Research Council (MC-A658-5QEB0); National Institute for Health Research (NIHR) Imperial College Biomedical Research Centre; British Heart Foundation (NH/17/1/32725, RG/19/6/34387, RE/18/4/34215); Academy of Medical Sciences (SGL015/1006); Mason Medical Research Trust grant; and the Engineering and Physical Sciences Research Council (EP/P001009/1).

\bibliographystyle{model2-names.bst}\biboptions{authoryear}
\bibliography{references.bib}

\begin{thebibliography}{46}
\expandafter\ifx\csname natexlab\endcsname\relax\def\natexlab#1{#1}\fi
\providecommand{\url}[1]{\texttt{#1}}
\providecommand{\href}[2]{#2}
\providecommand{\path}[1]{#1}
\providecommand{\DOIprefix}{doi:}
\providecommand{\ArXivprefix}{arXiv:}
\providecommand{\URLprefix}{URL: }
\providecommand{\Pubmedprefix}{pmid:}
\providecommand{\doi}[1]{\href{http://dx.doi.org/#1}{\path{#1}}}
\providecommand{\Pubmed}[1]{\href{pmid:#1}{\path{#1}}}
\providecommand{\bibinfo}[2]{#2}
\ifx\xfnm\relax \def\xfnm[#1]{\unskip,\space#1}\fi
%Type = Article
\bibitem[{Bai et~al.(2015)Bai, Shi, de~Marvao, Dawes, O’Regan, Cook and
  Rueckert}]{bai2015bi}
\bibinfo{author}{Bai, W.}, \bibinfo{author}{Shi, W.},
  \bibinfo{author}{de~Marvao, A.}, \bibinfo{author}{Dawes, T.J.},
  \bibinfo{author}{O’Regan, D.P.}, \bibinfo{author}{Cook, S.A.},
  \bibinfo{author}{Rueckert, D.}, \bibinfo{year}{2015}.
\newblock \bibinfo{title}{A bi-ventricular cardiac atlas built from 1000+ high
  resolution {MR} images of healthy subjects and an analysis of shape and
  motion}.
\newblock \bibinfo{journal}{Med Image Anal} \bibinfo{volume}{26},
  \bibinfo{pages}{133--145}.
%Type = Inproceedings
\bibitem[{Bao et~al.(2019)Bao, Lai, Ma, Zhang, Gao and Yang}]{bao2019depth}
\bibinfo{author}{Bao, W.}, \bibinfo{author}{Lai, W.S.}, \bibinfo{author}{Ma,
  C.}, \bibinfo{author}{Zhang, X.}, \bibinfo{author}{Gao, Z.},
  \bibinfo{author}{Yang, M.H.}, \bibinfo{year}{2019}.
\newblock \bibinfo{title}{Depth-aware video frame interpolation}, in:
  \bibinfo{booktitle}{Proceedings of the IEEE/CVF Conference on Computer Vision
  and Pattern Recognition}, pp. \bibinfo{pages}{3703--3712}.
%Type = Article
\bibitem[{Bello et~al.(2019)Bello, Dawes, Duan, Biffi, de~Marvao, Howard,
  Gibbs, Wilkins, Cook, Rueckert et~al.}]{bello2019deep}
\bibinfo{author}{Bello, G.A.}, \bibinfo{author}{Dawes, T.J.},
  \bibinfo{author}{Duan, J.}, \bibinfo{author}{Biffi, C.},
  \bibinfo{author}{de~Marvao, A.}, \bibinfo{author}{Howard, L.S.},
  \bibinfo{author}{Gibbs, J.S.R.}, \bibinfo{author}{Wilkins, M.R.},
  \bibinfo{author}{Cook, S.A.}, \bibinfo{author}{Rueckert, D.}, et~al.,
  \bibinfo{year}{2019}.
\newblock \bibinfo{title}{Deep-learning cardiac motion analysis for human
  survival prediction}.
\newblock \bibinfo{journal}{Nat Mach Intell} \bibinfo{volume}{1},
  \bibinfo{pages}{95--104}.
%Type = Article
\bibitem[{Bycroft et~al.(2018)Bycroft, Freeman, Petkova, Band, Elliott, Sharp,
  Motyer, Vukcevic, Delaneau, O’Connell et~al.}]{bycroft2018uk}
\bibinfo{author}{Bycroft, C.}, \bibinfo{author}{Freeman, C.},
  \bibinfo{author}{Petkova, D.}, \bibinfo{author}{Band, G.},
  \bibinfo{author}{Elliott, L.T.}, \bibinfo{author}{Sharp, K.},
  \bibinfo{author}{Motyer, A.}, \bibinfo{author}{Vukcevic, D.},
  \bibinfo{author}{Delaneau, O.}, \bibinfo{author}{O’Connell, J.}, et~al.,
  \bibinfo{year}{2018}.
\newblock \bibinfo{title}{The {UK} biobank resource with deep phenotyping and
  genomic data}.
\newblock \bibinfo{journal}{Nature} \bibinfo{volume}{562},
  \bibinfo{pages}{203--209}.
%Type = Article
\bibitem[{Cao et~al.(2020)Cao, Wu and Kr{\"a}henb{\"u}hl}]{cao2020lossless}
\bibinfo{author}{Cao, S.}, \bibinfo{author}{Wu, C.Y.},
  \bibinfo{author}{Kr{\"a}henb{\"u}hl, P.}, \bibinfo{year}{2020}.
\newblock \bibinfo{title}{Lossless image compression through super-resolution}.
\newblock \bibinfo{journal}{arXiv preprint arXiv:2004.02872} .
%Type = Article
\bibitem[{Chaudhari et~al.(2018)Chaudhari, Fang, Kogan, Wood, Stevens, Gibbons,
  Lee, Gold and Hargreaves}]{chaudhari2018super}
\bibinfo{author}{Chaudhari, A.S.}, \bibinfo{author}{Fang, Z.},
  \bibinfo{author}{Kogan, F.}, \bibinfo{author}{Wood, J.},
  \bibinfo{author}{Stevens, K.J.}, \bibinfo{author}{Gibbons, E.K.},
  \bibinfo{author}{Lee, J.H.}, \bibinfo{author}{Gold, G.E.},
  \bibinfo{author}{Hargreaves, B.A.}, \bibinfo{year}{2018}.
\newblock \bibinfo{title}{Super-resolution musculoskeletal {MRI} using deep
  learning}.
\newblock \bibinfo{journal}{Magn Reson Med} \bibinfo{volume}{80},
  \bibinfo{pages}{2139--2154}.
%Type = Inproceedings
\bibitem[{Chen et~al.(2018)Chen, Shi, Christodoulou, Xie, Zhou and
  Li}]{chen2018efficient}
\bibinfo{author}{Chen, Y.}, \bibinfo{author}{Shi, F.},
  \bibinfo{author}{Christodoulou, A.G.}, \bibinfo{author}{Xie, Y.},
  \bibinfo{author}{Zhou, Z.}, \bibinfo{author}{Li, D.}, \bibinfo{year}{2018}.
\newblock \bibinfo{title}{Efficient and accurate {MRI} super-resolution using a
  generative adversarial network and 3{D} multi-level densely connected
  network}, in: \bibinfo{booktitle}{International Conference on Medical Image
  Computing and Computer-Assisted Intervention},
  \bibinfo{organization}{Springer}. pp. \bibinfo{pages}{91--99}.
%Type = Inproceedings
\bibitem[{Cui et~al.(2014)Cui, Chang, Shan, Zhong and Chen}]{cui2014deep}
\bibinfo{author}{Cui, Z.}, \bibinfo{author}{Chang, H.}, \bibinfo{author}{Shan,
  S.}, \bibinfo{author}{Zhong, B.}, \bibinfo{author}{Chen, X.},
  \bibinfo{year}{2014}.
\newblock \bibinfo{title}{Deep network cascade for image super-resolution}, in:
  \bibinfo{booktitle}{European Conference on Computer Vision},
  \bibinfo{organization}{Springer}. pp. \bibinfo{pages}{49--64}.
%Type = Article
\bibitem[{Davatzikos et~al.(2003)Davatzikos, Tao and
  Shen}]{davatzikos2003hierarchical}
\bibinfo{author}{Davatzikos, C.}, \bibinfo{author}{Tao, X.},
  \bibinfo{author}{Shen, D.}, \bibinfo{year}{2003}.
\newblock \bibinfo{title}{Hierarchical active shape models, using the wavelet
  transform}.
\newblock \bibinfo{journal}{IEEE Trans Med Imaging} \bibinfo{volume}{22},
  \bibinfo{pages}{414--423}.
%Type = Article
\bibitem[{Delannoy et~al.(2020)Delannoy, Pham, Cazorla, Tor-D{\'\i}ez,
  Doll{\'e}, Meunier, Bednarek, Fablet, Passat and
  Rousseau}]{delannoy2020segsrgan}
\bibinfo{author}{Delannoy, Q.}, \bibinfo{author}{Pham, C.H.},
  \bibinfo{author}{Cazorla, C.}, \bibinfo{author}{Tor-D{\'\i}ez, C.},
  \bibinfo{author}{Doll{\'e}, G.}, \bibinfo{author}{Meunier, H.},
  \bibinfo{author}{Bednarek, N.}, \bibinfo{author}{Fablet, R.},
  \bibinfo{author}{Passat, N.}, \bibinfo{author}{Rousseau, F.},
  \bibinfo{year}{2020}.
\newblock \bibinfo{title}{{SegSRGAN}: Super-resolution and segmentation using
  generative adversarial networks—application to neonatal brain {MRI}}.
\newblock \bibinfo{journal}{Comput Biol Med} \bibinfo{volume}{120},
  \bibinfo{pages}{103755}.
%Type = Inproceedings
\bibitem[{Dong et~al.(2014)Dong, Loy, He and Tang}]{dong2014learning}
\bibinfo{author}{Dong, C.}, \bibinfo{author}{Loy, C.C.}, \bibinfo{author}{He,
  K.}, \bibinfo{author}{Tang, X.}, \bibinfo{year}{2014}.
\newblock \bibinfo{title}{Learning a deep convolutional network for image
  super-resolution}, in: \bibinfo{booktitle}{European conference on computer
  vision}, \bibinfo{organization}{Springer}. pp. \bibinfo{pages}{184--199}.
%Type = Article
\bibitem[{Dou et~al.(2018)Dou, Ouyang, Chen, Chen and
  Heng}]{dou2018unsupervised}
\bibinfo{author}{Dou, Q.}, \bibinfo{author}{Ouyang, C.}, \bibinfo{author}{Chen,
  C.}, \bibinfo{author}{Chen, H.}, \bibinfo{author}{Heng, P.A.},
  \bibinfo{year}{2018}.
\newblock \bibinfo{title}{Unsupervised cross-modality domain adaptation of
  convnets for biomedical image segmentations with adversarial loss}.
\newblock \bibinfo{journal}{arXiv preprint arXiv:1804.10916} .
%Type = Article
\bibitem[{{Duan} et~al.(2019){Duan}, {Bello}, {Schlemper}, {Bai}, {Dawes},
  {Biffi}, {de Marvao}, {Doumoud}, {O’Regan} and {Rueckert}}]{8624549}
\bibinfo{author}{{Duan}, J.}, \bibinfo{author}{{Bello}, G.},
  \bibinfo{author}{{Schlemper}, J.}, \bibinfo{author}{{Bai}, W.},
  \bibinfo{author}{{Dawes}, T.J.W.}, \bibinfo{author}{{Biffi}, C.},
  \bibinfo{author}{{de Marvao}, A.}, \bibinfo{author}{{Doumoud}, G.},
  \bibinfo{author}{{O’Regan}, D.P.}, \bibinfo{author}{{Rueckert}, D.},
  \bibinfo{year}{2019}.
\newblock \bibinfo{title}{Automatic 3{D} bi-ventricular segmentation of cardiac
  images by a shape-refined multi- task deep learning approach}.
\newblock \bibinfo{journal}{IEEE Trans Med Imaging} \bibinfo{volume}{38},
  \bibinfo{pages}{2151--2164}.
\newblock \DOIprefix\doi{10.1109/TMI.2019.2894322}.
%Type = Article
\bibitem[{Durall et~al.(2020)Durall, Chatzimichailidis, Labus and
  Keuper}]{durall2020combating}
\bibinfo{author}{Durall, R.}, \bibinfo{author}{Chatzimichailidis, A.},
  \bibinfo{author}{Labus, P.}, \bibinfo{author}{Keuper, J.},
  \bibinfo{year}{2020}.
\newblock \bibinfo{title}{Combating mode collapse in {GAN} training: An
  empirical analysis using {H}essian {E}igenvalues}.
\newblock \bibinfo{journal}{arXiv preprint arXiv:2012.09673} .
%Type = Inproceedings
\bibitem[{Ghifary et~al.(2016)Ghifary, Kleijn, Zhang, Balduzzi and
  Li}]{ghifary2016deep}
\bibinfo{author}{Ghifary, M.}, \bibinfo{author}{Kleijn, W.B.},
  \bibinfo{author}{Zhang, M.}, \bibinfo{author}{Balduzzi, D.},
  \bibinfo{author}{Li, W.}, \bibinfo{year}{2016}.
\newblock \bibinfo{title}{Deep reconstruction-classification networks for
  unsupervised domain adaptation}, in: \bibinfo{booktitle}{European Conference
  on Computer Vision}, \bibinfo{organization}{Springer}. pp.
  \bibinfo{pages}{597--613}.
%Type = Article
\bibitem[{Gholipour et~al.(2010)Gholipour, Estroff and
  Warfield}]{gholipour2010robust}
\bibinfo{author}{Gholipour, A.}, \bibinfo{author}{Estroff, J.A.},
  \bibinfo{author}{Warfield, S.K.}, \bibinfo{year}{2010}.
\newblock \bibinfo{title}{Robust super-resolution volume reconstruction from
  slice acquisitions: application to fetal brain {MRI}}.
\newblock \bibinfo{journal}{IEEE Trans Med Imaging} \bibinfo{volume}{29},
  \bibinfo{pages}{1739--1758}.
%Type = Article
\bibitem[{He et~al.(2015)He, Zhang, Ren and Sun}]{DBLP:journals/corr/HeZRS15}
\bibinfo{author}{He, K.}, \bibinfo{author}{Zhang, X.}, \bibinfo{author}{Ren,
  S.}, \bibinfo{author}{Sun, J.}, \bibinfo{year}{2015}.
\newblock \bibinfo{title}{Deep residual learning for image recognition}.
\newblock \bibinfo{journal}{arXiv preprint arXiv:1512.03385} .
%Type = Inproceedings
\bibitem[{Hoffman et~al.(2018)Hoffman, Tzeng, Park, Zhu, Isola, Saenko, Efros
  and Darrell}]{hoffman2018cycada}
\bibinfo{author}{Hoffman, J.}, \bibinfo{author}{Tzeng, E.},
  \bibinfo{author}{Park, T.}, \bibinfo{author}{Zhu, J.Y.},
  \bibinfo{author}{Isola, P.}, \bibinfo{author}{Saenko, K.},
  \bibinfo{author}{Efros, A.}, \bibinfo{author}{Darrell, T.},
  \bibinfo{year}{2018}.
\newblock \bibinfo{title}{Cycada: Cycle-consistent adversarial domain
  adaptation}, in: \bibinfo{booktitle}{International conference on machine
  learning}, \bibinfo{organization}{PMLR}. pp. \bibinfo{pages}{1989--1998}.
%Type = Inproceedings
\bibitem[{Huang et~al.(2018)Huang, Liu, Belongie and
  Kautz}]{huang2018multimodal}
\bibinfo{author}{Huang, X.}, \bibinfo{author}{Liu, M.Y.},
  \bibinfo{author}{Belongie, S.}, \bibinfo{author}{Kautz, J.},
  \bibinfo{year}{2018}.
\newblock \bibinfo{title}{Multimodal unsupervised image-to-image translation},
  in: \bibinfo{booktitle}{Proceedings of the European conference on computer
  vision (ECCV)}, pp. \bibinfo{pages}{172--189}.
%Type = Article
\bibitem[{Kingma and Ba(2014)}]{kingma2014adam}
\bibinfo{author}{Kingma, D.P.}, \bibinfo{author}{Ba, J.}, \bibinfo{year}{2014}.
\newblock \bibinfo{title}{Adam: A method for stochastic optimization}.
\newblock \bibinfo{journal}{arXiv preprint arXiv:1412.6980} .
%Type = Article
\bibitem[{Kingma and Welling(2013)}]{kingma2013auto}
\bibinfo{author}{Kingma, D.P.}, \bibinfo{author}{Welling, M.},
  \bibinfo{year}{2013}.
\newblock \bibinfo{title}{Auto-encoding variational {B}ayes}.
\newblock \bibinfo{journal}{arXiv preprint arXiv:1312.6114} .
%Type = Article
\bibitem[{Kingma and Welling(2019)}]{kingma2019introduction}
\bibinfo{author}{Kingma, D.P.}, \bibinfo{author}{Welling, M.},
  \bibinfo{year}{2019}.
\newblock \bibinfo{title}{An introduction to variational autoencoders}.
\newblock \bibinfo{journal}{arXiv preprint arXiv:1906.02691} .
%Type = Inproceedings
\bibitem[{Ledig et~al.(2017)Ledig, Theis, Husz{\'a}r, Caballero, Cunningham,
  Acosta, Aitken, Tejani, Totz, Wang et~al.}]{ledig2017photo}
\bibinfo{author}{Ledig, C.}, \bibinfo{author}{Theis, L.},
  \bibinfo{author}{Husz{\'a}r, F.}, \bibinfo{author}{Caballero, J.},
  \bibinfo{author}{Cunningham, A.}, \bibinfo{author}{Acosta, A.},
  \bibinfo{author}{Aitken, A.}, \bibinfo{author}{Tejani, A.},
  \bibinfo{author}{Totz, J.}, \bibinfo{author}{Wang, Z.}, et~al.,
  \bibinfo{year}{2017}.
\newblock \bibinfo{title}{Photo-realistic single image super-resolution using a
  generative adversarial network}, in: \bibinfo{booktitle}{Proceedings of the
  IEEE conference on computer vision and pattern recognition}, pp.
  \bibinfo{pages}{4681--4690}.
%Type = Article
\bibitem[{Moghari et~al.(2018)Moghari, Barthur, Amaral, Geva and
  Powell}]{moghari2018free}
\bibinfo{author}{Moghari, M.H.}, \bibinfo{author}{Barthur, A.},
  \bibinfo{author}{Amaral, M.E.}, \bibinfo{author}{Geva, T.},
  \bibinfo{author}{Powell, A.J.}, \bibinfo{year}{2018}.
\newblock \bibinfo{title}{Free-breathing whole-heart 3{D} cine magnetic
  resonance imaging with prospective respiratory motion compensation}.
\newblock \bibinfo{journal}{Magn Reson Med} \bibinfo{volume}{80},
  \bibinfo{pages}{181--189}.
%Type = Article
\bibitem[{Oktay et~al.(2017)Oktay, Ferrante, Kamnitsas, Heinrich, Bai,
  Caballero, Cook, De~Marvao, Dawes, O‘Regan et~al.}]{oktay2017anatomically}
\bibinfo{author}{Oktay, O.}, \bibinfo{author}{Ferrante, E.},
  \bibinfo{author}{Kamnitsas, K.}, \bibinfo{author}{Heinrich, M.},
  \bibinfo{author}{Bai, W.}, \bibinfo{author}{Caballero, J.},
  \bibinfo{author}{Cook, S.A.}, \bibinfo{author}{De~Marvao, A.},
  \bibinfo{author}{Dawes, T.}, \bibinfo{author}{O‘Regan, D.P.}, et~al.,
  \bibinfo{year}{2017}.
\newblock \bibinfo{title}{Anatomically constrained neural networks ({ACNN}s):
  application to cardiac image enhancement and segmentation}.
\newblock \bibinfo{journal}{IEEE Trans Med Imaging} \bibinfo{volume}{37},
  \bibinfo{pages}{384--395}.
%Type = Article
\bibitem[{Perone et~al.(2019)Perone, Ballester, Barros and
  Cohen-Adad}]{perone2019unsupervised}
\bibinfo{author}{Perone, C.S.}, \bibinfo{author}{Ballester, P.},
  \bibinfo{author}{Barros, R.C.}, \bibinfo{author}{Cohen-Adad, J.},
  \bibinfo{year}{2019}.
\newblock \bibinfo{title}{Unsupervised domain adaptation for medical imaging
  segmentation with self-ensembling}.
\newblock \bibinfo{journal}{NeuroImage} \bibinfo{volume}{194},
  \bibinfo{pages}{1--11}.
%Type = Article
\bibitem[{Petersen et~al.(2015)Petersen, Matthews, Francis, Robson, Zemrak,
  Boubertakh, Young, Hudson, Weale, Garratt et~al.}]{petersen2015uk}
\bibinfo{author}{Petersen, S.E.}, \bibinfo{author}{Matthews, P.M.},
  \bibinfo{author}{Francis, J.M.}, \bibinfo{author}{Robson, M.D.},
  \bibinfo{author}{Zemrak, F.}, \bibinfo{author}{Boubertakh, R.},
  \bibinfo{author}{Young, A.A.}, \bibinfo{author}{Hudson, S.},
  \bibinfo{author}{Weale, P.}, \bibinfo{author}{Garratt, S.}, et~al.,
  \bibinfo{year}{2015}.
\newblock \bibinfo{title}{{UK} {B}iobank’s cardiovascular magnetic resonance
  protocol}.
\newblock \bibinfo{journal}{J Cardiovasc Magn Reson} \bibinfo{volume}{18},
  \bibinfo{pages}{8}.
%Type = Inproceedings
\bibitem[{Rasti et~al.(2016)Rasti, Uiboupin, Escalera and
  Anbarjafari}]{rasti2016convolutional}
\bibinfo{author}{Rasti, P.}, \bibinfo{author}{Uiboupin, T.},
  \bibinfo{author}{Escalera, S.}, \bibinfo{author}{Anbarjafari, G.},
  \bibinfo{year}{2016}.
\newblock \bibinfo{title}{Convolutional neural network super resolution for
  face recognition in surveillance monitoring}, in:
  \bibinfo{booktitle}{International conference on articulated motion and
  deformable objects}, \bibinfo{organization}{Springer}. pp.
  \bibinfo{pages}{175--184}.
%Type = Article
\bibitem[{Rezende et~al.(2014)Rezende, Mohamed and
  Wierstra}]{rezende2014stochastic}
\bibinfo{author}{Rezende, D.J.}, \bibinfo{author}{Mohamed, S.},
  \bibinfo{author}{Wierstra, D.}, \bibinfo{year}{2014}.
\newblock \bibinfo{title}{Stochastic backpropagation and approximate inference
  in deep generative models (2014)}.
\newblock \bibinfo{journal}{arXiv preprint arXiv:1401.4082} .
%Type = Inproceedings
\bibitem[{Ronneberger et~al.(2015)Ronneberger, Fischer and
  Brox}]{ronneberger2015u}
\bibinfo{author}{Ronneberger, O.}, \bibinfo{author}{Fischer, P.},
  \bibinfo{author}{Brox, T.}, \bibinfo{year}{2015}.
\newblock \bibinfo{title}{U-net: Convolutional networks for biomedical image
  segmentation}, in: \bibinfo{booktitle}{International Conference on Medical
  image computing and computer-assisted intervention},
  \bibinfo{organization}{Springer}. pp. \bibinfo{pages}{234--241}.
%Type = Inproceedings
\bibitem[{Sabour et~al.(2017a)Sabour, Frosst and Hinton}]{46351}
\bibinfo{author}{Sabour, S.}, \bibinfo{author}{Frosst, N.},
  \bibinfo{author}{Hinton, G.}, \bibinfo{year}{2017}a.
\newblock \bibinfo{title}{Dynamic routing between capsules}.
%Type = Article
\bibitem[{Sabour et~al.(2017b)Sabour, Frosst and Hinton}]{sabour2017dynamic}
\bibinfo{author}{Sabour, S.}, \bibinfo{author}{Frosst, N.},
  \bibinfo{author}{Hinton, G.E.}, \bibinfo{year}{2017}b.
\newblock \bibinfo{title}{Dynamic routing between capsules}.
\newblock \bibinfo{journal}{arXiv preprint arXiv:1710.09829} .
%Type = Article
\bibitem[{S{\'a}nchez and Vilaplana(2018)}]{sanchez2018brain}
\bibinfo{author}{S{\'a}nchez, I.}, \bibinfo{author}{Vilaplana, V.},
  \bibinfo{year}{2018}.
\newblock \bibinfo{title}{Brain {MRI} super-resolution using 3{D} generative
  adversarial networks}.
\newblock \bibinfo{journal}{arXiv preprint arXiv:1812.11440} .
%Type = Article
\bibitem[{Savioli et~al.(2021a)Savioli, de~Marvao, Bai, Wang, Cook, Chin,
  Rueckert and {O’Regan}}]{github}
\bibinfo{author}{Savioli, N.}, \bibinfo{author}{de~Marvao, A.},
  \bibinfo{author}{Bai, W.}, \bibinfo{author}{Wang, S.}, \bibinfo{author}{Cook,
  S.A.}, \bibinfo{author}{Chin, C.W.L.}, \bibinfo{author}{Rueckert, D.},
  \bibinfo{author}{{O’Regan}, D.P.}, \bibinfo{year}{2021}a.
\newblock \bibinfo{title}{{ImperialCollegeLondon/Gemini-GAN}}.
\newblock \bibinfo{journal}{Zenodo} \DOIprefix\doi{10.5281/zenodo.5005942}.
%Type = Article
\bibitem[{Savioli et~al.(2021b)Savioli, de~Marvao and {O’Regan}}]{data}
\bibinfo{author}{Savioli, N.}, \bibinfo{author}{de~Marvao, A.},
  \bibinfo{author}{{O’Regan}, D.P.}, \bibinfo{year}{2021}b.
\newblock \bibinfo{title}{Cardiac super-resolution label maps}.
\newblock \bibinfo{journal}{Mendeley Data} \bibinfo{volume}{V1}.
\newblock \DOIprefix\doi{10.17632/pw87p286yx.1}.
%Type = Article
\bibitem[{Schafer et~al.(2017)Schafer, De~Marvao, Adami, Fiedler, Ng, Khin,
  Rackham, Van~Heesch, Pua, Kui et~al.}]{schafer2017titin}
\bibinfo{author}{Schafer, S.}, \bibinfo{author}{De~Marvao, A.},
  \bibinfo{author}{Adami, E.}, \bibinfo{author}{Fiedler, L.R.},
  \bibinfo{author}{Ng, B.}, \bibinfo{author}{Khin, E.},
  \bibinfo{author}{Rackham, O.J.}, \bibinfo{author}{Van~Heesch, S.},
  \bibinfo{author}{Pua, C.J.}, \bibinfo{author}{Kui, M.}, et~al.,
  \bibinfo{year}{2017}.
\newblock \bibinfo{title}{Titin-truncating variants affect heart function in
  disease cohorts and the general population}.
\newblock \bibinfo{journal}{Nat Genet} \bibinfo{volume}{49},
  \bibinfo{pages}{46--53}.
%Type = Article
\bibitem[{Shui et~al.(2020)Shui, Chen, Wen, Zhou, Gagn{\'e} and
  Wang}]{DBLP:journals/corr/abs-2007-15567}
\bibinfo{author}{Shui, C.}, \bibinfo{author}{Chen, Q.}, \bibinfo{author}{Wen,
  J.}, \bibinfo{author}{Zhou, F.}, \bibinfo{author}{Gagn{\'e}, C.},
  \bibinfo{author}{Wang, B.}, \bibinfo{year}{2020}.
\newblock \bibinfo{title}{Beyond {H}-divergence: {D}omain adaptation theory
  with {J}ensen-{S}hannon divergence}.
\newblock \bibinfo{journal}{arXiv preprint arXiv:2007.15567} .
%Type = Article
\bibitem[{Tzeng et~al.(2014)Tzeng, Hoffman, Zhang, Saenko and
  Darrell}]{tzeng2014deep}
\bibinfo{author}{Tzeng, E.}, \bibinfo{author}{Hoffman, J.},
  \bibinfo{author}{Zhang, N.}, \bibinfo{author}{Saenko, K.},
  \bibinfo{author}{Darrell, T.}, \bibinfo{year}{2014}.
\newblock \bibinfo{title}{Deep domain confusion: Maximizing for domain
  invariance}.
\newblock \bibinfo{journal}{arXiv preprint arXiv:1412.3474} .
%Type = Article
\bibitem[{Wang and Deng(2018)}]{wang2018deep}
\bibinfo{author}{Wang, M.}, \bibinfo{author}{Deng, W.}, \bibinfo{year}{2018}.
\newblock \bibinfo{title}{Deep visual domain adaptation: A survey}.
\newblock \bibinfo{journal}{Neurocomputing} \bibinfo{volume}{312},
  \bibinfo{pages}{135--153}.
%Type = Article
\bibitem[{Wang et~al.(2004)Wang, Bovik, Sheikh and Simoncelli}]{wang2004image}
\bibinfo{author}{Wang, Z.}, \bibinfo{author}{Bovik, A.C.},
  \bibinfo{author}{Sheikh, H.R.}, \bibinfo{author}{Simoncelli, E.P.},
  \bibinfo{year}{2004}.
\newblock \bibinfo{title}{Image quality assessment: from error visibility to
  structural similarity}.
\newblock \bibinfo{journal}{IEEE Trans Image Process} \bibinfo{volume}{13},
  \bibinfo{pages}{600--612}.
%Type = Article
\bibitem[{Woodbridge et~al.(2013)Woodbridge, Fagiolo and
  O’Regan}]{woodbridge2013mridb}
\bibinfo{author}{Woodbridge, M.}, \bibinfo{author}{Fagiolo, G.},
  \bibinfo{author}{O’Regan, D.P.}, \bibinfo{year}{2013}.
\newblock \bibinfo{title}{{MRI}db: medical image management for biobank
  research}.
\newblock \bibinfo{journal}{J Digit Imaging} \bibinfo{volume}{26},
  \bibinfo{pages}{886--890}.
%Type = Article
\bibitem[{Xia et~al.(2021)Xia, Ravikumar, Greenwood, Neubauer, Petersen and
  Frangi}]{xia2021super}
\bibinfo{author}{Xia, Y.}, \bibinfo{author}{Ravikumar, N.},
  \bibinfo{author}{Greenwood, J.P.}, \bibinfo{author}{Neubauer, S.},
  \bibinfo{author}{Petersen, S.E.}, \bibinfo{author}{Frangi, A.F.},
  \bibinfo{year}{2021}.
\newblock \bibinfo{title}{Super-resolution of cardiac {MR} cine imaging using
  conditional {GAN}s and unsupervised transfer learning}.
\newblock \bibinfo{journal}{Med Image Anal} \bibinfo{volume}{71},
  \bibinfo{pages}{102037}.
%Type = Inproceedings
\bibitem[{Yang et~al.(2020)Yang, An, Wang, Zhu, Yan and Huang}]{yang2020label}
\bibinfo{author}{Yang, J.}, \bibinfo{author}{An, W.}, \bibinfo{author}{Wang,
  S.}, \bibinfo{author}{Zhu, X.}, \bibinfo{author}{Yan, C.},
  \bibinfo{author}{Huang, J.}, \bibinfo{year}{2020}.
\newblock \bibinfo{title}{Label-driven reconstruction for domain adaptation in
  semantic segmentation}, in: \bibinfo{booktitle}{European Conference on
  Computer Vision}, \bibinfo{organization}{Springer}. pp.
  \bibinfo{pages}{480--498}.
%Type = Inproceedings
\bibitem[{Zhu et~al.(2017a)Zhu, Park, Isola and Efros}]{zhu2017unpaired}
\bibinfo{author}{Zhu, J.Y.}, \bibinfo{author}{Park, T.},
  \bibinfo{author}{Isola, P.}, \bibinfo{author}{Efros, A.A.},
  \bibinfo{year}{2017}a.
\newblock \bibinfo{title}{Unpaired image-to-image translation using
  cycle-consistent adversarial networks}, in: \bibinfo{booktitle}{Proceedings
  of the IEEE international conference on computer vision}, pp.
  \bibinfo{pages}{2223--2232}.
%Type = Article
\bibitem[{Zhu et~al.(2017b)Zhu, Zhang, Pathak, Darrell, Efros, Wang and
  Shechtman}]{zhu2017toward}
\bibinfo{author}{Zhu, J.Y.}, \bibinfo{author}{Zhang, R.},
  \bibinfo{author}{Pathak, D.}, \bibinfo{author}{Darrell, T.},
  \bibinfo{author}{Efros, A.A.}, \bibinfo{author}{Wang, O.},
  \bibinfo{author}{Shechtman, E.}, \bibinfo{year}{2017}b.
\newblock \bibinfo{title}{Toward multimodal image-to-image translation}.
\newblock \bibinfo{journal}{arXiv preprint arXiv:1711.11586} .
%Type = Inproceedings
\bibitem[{Zhu et~al.(2014)Zhu, Zhang and Yuille}]{zhu2014single}
\bibinfo{author}{Zhu, Y.}, \bibinfo{author}{Zhang, Y.},
  \bibinfo{author}{Yuille, A.L.}, \bibinfo{year}{2014}.
\newblock \bibinfo{title}{Single image super-resolution using deformable
  patches}, in: \bibinfo{booktitle}{Proceedings of the IEEE Conference on
  Computer Vision and Pattern Recognition}, pp. \bibinfo{pages}{2917--2924}.

\end{thebibliography}

\appendix

\section{Theorem 1 } \label{appendix:Theorem1}

In this appendix section, we are going to demonstrate how the VAE loss is given by the combination of Evidence Lower Bound (ELBO) and Kullback-Leibler (KL) divergence and how derivative a gradient forms from it. Indeed, the latent variable $z$ is part of the variational target generator model but we can't directly observe it. In this specific case the direct graphical model is represented by a joint distribution $p_{\theta}(x;z)$ over both LR input data target volume $x$ and latent variable $z$. Where, the marginal distribution $p_{\theta}(x)$, over the observed variable $x$, is given by: 

\begin{equation} \label{eq:9}
\begin{split}
p_{\theta}(x) = \int p_{\theta} (x,z) dz
\end{split}
\end{equation}

The $\theta$ are the parameters that minimize the integral. However, this marginal distribution $p_{\theta}(x)$ can be complicated and need to be approximate through a factoring process:

\begin{equation} \label{eq:10}
p_{\theta}(x,z) = p_{\theta}(x)p_{\theta} (x|z)
\end{equation}

The distribution $p(z)$ is also called prior distribution over latent $z$ variables. The challenge is due to the intractability of $p_{\theta}(x,z)$, where no analytic solution is given. To solve this intractability, a parametric inference model $q_{\phi}(z|x)$ is then introduce trough an encoder network (i.e., $G_{\phi}$) that optimize the variational parameters $\phi$ as $q_{\phi} (z|x) \approx p_{\theta} (z|x)$. In the variational inference, this is called lower bound objective (ELBO) optimisation problem \citep{kingma2019introduction}. Then, for any type of parametric inference model $q_{\phi}(z|x)$, is given: 

\begin{equation} \label{eq:12}
\begin{split}
log (p_{\theta} (x)) = \mathbb{E}_{q_{\phi}(z|x)} [log p_{\theta} (x)] = \\
=  \mathbb{E}_{q_{\phi}(z|x)} [log (\frac{p_{\theta}(x,z)}{p_{\theta}(z|x)})] \\
=  \mathbb{E}_{q_{\phi}(z|x)}[log (\frac{p_{\theta}(x,z)}{q_{\phi}(z|x)} \frac{q_{\phi}(z|x)}{p_{\phi}(z|x)})] \\
= \mathbb{E}_{q_{\phi}(z|x)}[log(\frac{p_{\theta}(x,z)}{q_{\phi}(z|x)})] + \mathbb{E}_{q_{\phi}(z|x)}[log(\frac{q_{\phi}(z|x)}{p_{\theta}(z|x)}))]
\end{split}
\end{equation}

The first term, of the last equation line, is the ELBO (i.e., the final variational loss $Loss_{\theta,\phi}(x)$) and is always lower bound for $log (p_{\theta}(x))$ (i.e., due to the positive value of the second term). The ELBO is obtained by substitute the [Eq. \ref{eq:10}] on  [Eq. \ref{eq:12}] (i.e., first term of last equation line).

\begin{equation} \label{eq:13}
Loss_{\theta,\phi}(x) = \mathbb{E}_{q_{\phi}(z|x)}[log(\frac{p_{\theta}(x)p_{\theta}(x|z)} {q_{\phi}(z|x)})] 
\end{equation}

\begin{equation} \label{eq:14}
\begin{split}
Loss_{\theta,\phi}(x) = log p_{\theta} (x) -  D_{KL}  (q_{\phi} (z|x) || p_{\theta} (z|x)) \leq log p_{\theta}(x)
\end{split}
\end{equation}

The second term is the Kullback-Leibler (KL) divergence among the parametric inference model $q_{\phi}(z|x)$ and the marginal distribution $p_{\theta}(z|x)$ (i.e., also positive):

\begin{equation} \label{eq:15}
 D_{KL}  (q_{\phi} (z|x) || p_{\theta} (z|x))  = \mathbb{E}_{q_{\phi}(z|x)}[log(\frac{q_{\phi}(z|x)}{p_{\theta}(z|x)})] \geq 0
\end{equation}

Finally, we estimate the gradient for both parameters $\phi$ and $\theta$. In particular for $\theta$ we have: 

\begin{equation} \label{eq:16}
\begin{split}
\nabla_{\theta} Loss_{\theta,\phi} (x) = \nabla_{\theta} \mathbb{E}_{q_{\phi(z|x)}}[log(p_{\theta}(x,z))-log( q_{\phi}(z|x))] \\ 
=\mathbb{E}_{q_{\phi}(z|x)}[\nabla_{\theta}(log(p_{\theta}(x,z)) - log(q_{\phi}(z|x))] \\
\simeq \nabla_{\theta}(log(p_{\theta}(x,z))-log( q_{\phi}(z|x))) = \nabla_{\theta} (log p_{\theta}(x,z))
\end{split}
\end{equation}

Where the last term of this equation represents the Monte Carlo estimator for $z$ latent variable, randomly sampled from the parametric inference model $q_{\phi}(z|x)$ \citep{kingma2013auto}. However, for $\phi$ parameters there is no valid computational operation between expectation and gradient. Nonetheless, the $G_{{1}_{\phi}}$ encoder is fully differentiable and the latent variable $z$ used continued. Therefore, is calculate the gradient for $\phi$ variable with a reparameterization step \citep{rezende2014stochastic}. The reparameterization expresses the random variable $z$ in terms of another variable $k$ within a differentiable invertible transformation function $f(\cdot)$. The random noise of the reparametrization variable $k$ is sampled from a univariate Gaussian distribution $ \mathcal{N}(0,I)$.

\begin{equation} \label{eq:17}
z = f(k,\phi,x)
\end{equation}

The random distribution variable $k$ is independent both of the input variable $x$ and the parameter $\phi$. In this case the commutation operation between expectation and gradient  (i.e., both for the parameters $\theta$ and $\phi$) is guaranteed. Then, the new Monte Carlo estimator with $z = f(k,\phi,x)$ and random noise sample from $k \sim p(k)$, is given by the following equation:

\begin{equation} \label{eq:18}
\begin{split}
\mathbb{E}_{q_{\phi(z|x)}}[Loss_{\theta,\phi} (z)] = \mathbb{E}_{p(k)}[Loss_{\theta,\phi} (z)] \\
= \nabla_{\phi} \mathbb{E}_{q_{\phi(z|x)}}[Loss_{\theta,\phi} (z)] \\
=\mathbb{E}_{p(k)} [\nabla_{\phi} Loss_{\theta,\phi} (z)] \\
\simeq \nabla_{\phi} Loss_{\theta,\phi} (z)
\end{split}
\end{equation}

After the reparameterization trick: 

\begin{equation} \label{eq:19}
\begin{split}
k \sim  \mathcal{N}(0,I) \\
(\mu_{t}, log \sigma_{t}) = Encoding_{{G_{t}}_{\phi}}(x) \\
z =  \mu_{t} + \sigma_{t} \odot k
\end{split}
\end{equation}

Given the $q_{\phi}(z|x)$ is Gaussian and is approximate by $ Encoding_{{G_{x}}_{\phi}}$, the final inference is given by:

\begin{equation} \label{eq:8}
\begin{split}
(\mu_{x}, log (\sigma_{x})) = Encoding_{{G_{x}}_{\phi}}(I_{x}) \\
q_{\phi}(z_{x}|I_{i})_{x} = \mathcal{N}(z_{x};\mu_{x},diag(\sigma_{x})) \\
\end{split}
\end{equation}

Where $I_{x}$ indicated a general input image, $Encoding_{{G_{x}}_{\phi}}$ a generic encoding network, $z_{x}$ a generic latent variables of $Encoding_{{G_{x}}_{\phi}}$, and $\mu_{x}$, $\sigma_{x}$ the generic mean and standard deviation (i.e., predicted by $Encoding_{{G_{x}}_{\phi}}$) of the Gaussian distribution $\mathcal{N}(z_{x};\mu_{x},diag(\sigma_{x}))$, $k$ is the index of the sum operation. We can then conclude

\begin{equation} \label{eq:80}
\begin{split}
log(q_{\phi}(z_{x}|I_{x})_{x}) = \sum_{k} log(\mathcal{N}(z_{{x}_{k}},0,1)) - log (\sigma_{{x}_{k}}) \\
= \sum_{k} \frac{1}{2} (log(\sigma_{x_{k}}) - \mu^{2}_{x_{k}} - \sigma^{2}_{x_{k}} + 1)
\end{split}
\end{equation}

\section{Theorem 2} \label{appendix:Theorem2}

In this appendix section, we are going to demonstrate that if the variational generator $G_{{t}_{\phi}}$ and its discriminator $D_{t}$ are enough numerically stable, at each step iteration, the discriminator $D_{t}$ reach its optimum where the target output probability $p^{*}_{t}(x)$ of $I^{*}_{t}$  converges to the target data $I_{t}$ UKDHP distribution probability $p_{t}(x)$. 
Then, if we set $ K(p^{*}_{t}(x),D_{t})$ a function of the ``fake'' target probability distribution generate by $G_{{t}_{\phi}}$, the function $K(p^{*}_{t}(x),D_{t})$ shows a convexity in this ``fake'' probability distribution $p^{*}_{t}(x)$. 

To prove this, we let:

\begin{equation} \label{eq:22}
\begin{split}
K(p^{*}_{t}(x),D_{t}) = sup_{\alpha} K_{\alpha} (p^{*}_{t}(x),D_{t})
\end{split}
\end{equation}

Where the function $K_{\alpha}(p^{*}_{t}(x),D_{t})$ are convex on some convex domain $S$. We suppose that at particular point $p^{*}_{t}(x)$ in S,

\begin{equation} \label{eq:23}
\begin{split}
\beta = arg sup_{\alpha}  K_{\alpha}(p^{*}_{t}(x),D_{t}) 
\end{split}
\end{equation}

Then, 

\begin{equation} \label{eq:24}
\begin{split}
K(p^{*}_{t}(x),D_{t}) = K_{\beta} (p^{*}_{t}(x),D_{t})
\end{split}
\end{equation}

Let m be any subgradient of  $ K_{\beta} (p^{*}_{t}(x),D_{t})$, for definition of subgradient,

\begin{equation} \label{eq:25}
\begin{split}
K_{\beta} (p^{*}_{t}(y),D_{t}) \geq K_{\beta} (p^{*}_{t}(x),D_{t}) + m^{T} (y-x)
\end{split}
\end{equation}

Since $K(p^{*}_{t}(y),D_{t}) \geq  K_{\beta} (p^{*}_{t}(x),D_{t})$ for $y \in S$,

\begin{equation} \label{eq:26}
\begin{split}
K(p^{*}_{t}(y),D_{3}) \geq K(p^{*}_{t}(x),D_{3}) + m^{T} (y-x)
\end{split}
\end{equation}

Thus, we have $m \in \partial K(p^{*}_{t}(x),D_{t})$, but also for any subgradient  $\partial K_{\beta} (p^{*}_{t}(x),D_{t}) \in \partial  K(p^{*}_{t}(x),D_{t})$.
This is equivalent to calculate any update of Stochastic Gradient Descent
(SDG) for $p^{*}_{t}(x)$ at the optimal discriminator value $D^{optim}_{t}(x)$,  given the variational generator $G_{t\phi}$, to observed that any small update of generative ``fake'' distribution $p^{*}_{t}(x)$ and UKDHP data ``real'' distribution $p_{t}(x)$ converges to the data ``real'' distribution  $p^{*}_{t}(x)$; as $sup_{D^{optim}_{t}(x)} K(p^{*}_{t}(x),D^{optim}_{t}(x))$ is convex in $p^{*}_{t}(x)$, $p^{*}_{t} \xrightarrow{} p_{t}(x)$, concluding the proof.

%\section*{Supplementary Material}
%Supplementary material that may be helpful in the review process should
%be prepared and provided as a separate electronic file. That file can
%then be transformed into PDF format and submitted along with the
%manuscript and graphic files to the appropriate editorial office.

\end{document}